\documentclass[journal,twoside,web]{ieeecolor}
\usepackage{generic}
\usepackage{cite}
\usepackage{amsmath,amssymb,amsfonts}
\usepackage{algorithmic}
\usepackage{graphicx}
\usepackage{textcomp}
\usepackage{hyperref}
\usepackage{eqnarray}
\usepackage{multirow}
\usepackage{pifont}
\usepackage{todonotes}
\usepackage{lscape}
\usepackage{subfig}
\usepackage{amsmath}
\usepackage{rotating}
\usepackage{lineno}
\usepackage{times}
\usepackage{epsfig}
\usepackage{multirow}
\usepackage{pifont}

\def\BibTeX{{\rm B\kern-.05em{\sc i\kern-.025em b}\kern-.08em
    T\kern-.1667em\lower.7ex\hbox{E}\kern-.125emX}}
\begin{document}
\title{M$^3$Lung-Sys: A Deep Learning System for Multi-Class Lung Pneumonia Screening from CT Imaging}
\author{Xuelin Qian, Huazhu Fu, Weiya Shi, Tao Chen, Yanwei Fu, Fei Shan, and Xiangyang Xue
\thanks{Xuelin Qian and Xiangyang Xue are with Shanghai Key Lab of Intelligent Information Processing, School of Computer Science, Fudan University, Shanghai 200433, China. E-mail: \{xlqian15,xyxue\}@fudan.edu.cn.}
\thanks{Huazhu Fu is with Inception Institute of Artificial Intelligence, Abu Dhabi, UAE. E-mail: hzfu@ieee.org.}
\thanks{Weiya Shi and Fei Shan are with Department of Radiology, Shanghai Public Health Clinical Center, Fudan University, Shanghai 201508, China. E-mail: \{shiweiya,shanfei\}@shphc.org.cn.}
\thanks{Tao Chen is with School of Information Science and Technology, Fudan University, Shanghai 200433, China. E-mail: eetchen@fudan.edu.cn.}
\thanks{Yanwei Fu is with Shanghai Key Lab of Intelligent Information Processing, School of Big Data, Fudan University, Shanghai 200433, China. E-mail: yanweifu@fudan.edu.cn.}}
\maketitle

\begin{abstract}
To counter the outbreak of COVID-19, the accurate diagnosis of suspected cases plays a crucial role in timely quarantine, medical treatment, and preventing the spread of the pandemic. 
Considering the limited training cases and resources (\emph{e.g}, time and budget), we propose a Multi-task Multi-slice Deep Learning System (M$^3$Lung-Sys) for multi-class lung pneumonia screening from CT imaging, which only consists of two 2D CNN networks, \emph{i.e.}, slice- and patient-level classification networks. 
The former aims to seek the feature representations from abundant CT slices instead of limited CT volumes, and for the overall pneumonia screening, the latter one could recover the temporal information by feature refinement and aggregation between different slices. 
In addition to distinguish COVID-19 from Healthy, H1N1, and CAP cases, our M$^3$Lung-Sys also be able to locate the areas of relevant lesions, without any pixel-level annotation.
To further demonstrate the effectiveness of our model, we conduct extensive experiments on a chest CT imaging dataset with a total of 734 patients (251 healthy people, 245 COVID-19 patients, 105 H1N1 patients, and 133 CAP patients). 
The quantitative results with plenty of metrics indicate the superiority of our proposed model on both slice- and patient-level classification tasks.
More importantly, the generated lesion location maps make our system interpretable and more valuable to clinicians.
\end{abstract}


\section{Introduction}
\label{introduction}

Coronavirus disease 2019 (COVID-19), caused by a novel coronavirus (SARS-CoV-2, previously known as 2019-nCoV), is highly contagious and has become increasingly prevalent worldwide. 
The disease may lead to acute respiratory distress or multiple organ failure in severe cases~\cite{adhikari2020epidemiology,chan2020familial}. 
As of June 28th, 2020, $495,760$ of $9,843,073$ confirmed cases across countries have led to death, according to WHO statistics. 
Thus, how to accurately and efficiently diagnose COVID-19 is of vital importance not only for the timely treatment of patients, but also for the distribution and management of hospital resources during the outbreak.

The standard diagnostic method being used is real-time polymerase chain reaction (RT-PCR), which detects viral nucleotides from specimens obtained by oropharyngeal swab, nasopharyngeal swab, bronchoalveolar lavage, or tracheal aspirate~\cite{Interim_Guidelines}. 
Early reports of RT-PCR sensitivity vary considerably, ranging from 42\% to 71\%, and an initially negative RT-PCR result may convert into COVID-19 after up to four days~\cite{ai2020correlation}.
Recent studies have shown that typical Computed Tomography (CT) findings of COVID-19 include bilateral pulmonary parenchymal groundglass and consolidative pulmonary opacities, with a peripheral lung distribution~\cite{chung2020ct,pan2020imaging}. 
In contrast to RT-PCR, chest CT scans have demonstrated about 56$\sim$98\% sensitivity in detecting COVID-19 at initial manifestation and can be helpful in rectifying false negatives obtained from RT-PCR during early stages of disease development~\cite{fang2020sensitivity,kanne2020essentials}. 

\begin{figure}
\begin{centering}
\includegraphics[width=1\linewidth]{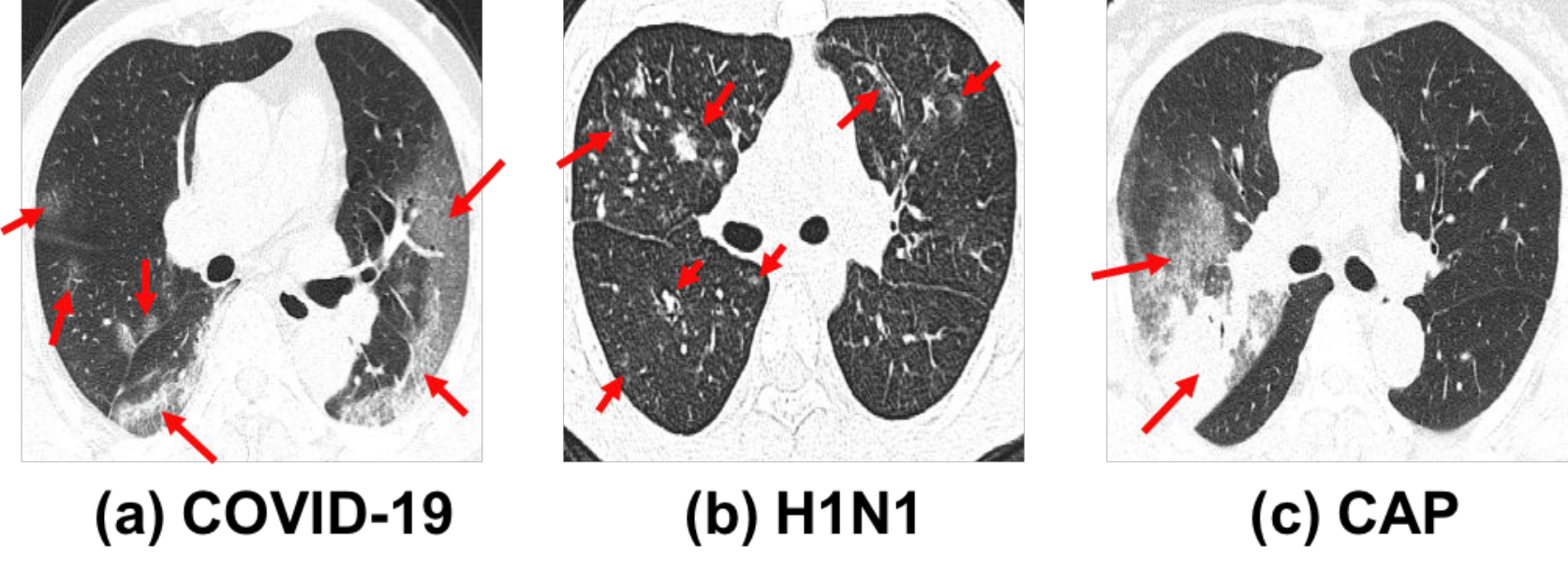} 
\caption{\label{fig:sample} 
Typical images of COVID-19, H1N1 and CAP. The red arrows indicate the locations of different lesions, which are marked by clinical experts. 
(a) COVID-19: CT shows ground glass opacity (GGO) with consolidation and crazy-paving sign distributed mainly along subpleural lungs. 
(b) H1N1(influenza A(H1N1)pdm09): the consolidation and small centrilobular nodules mainly locate at bronchovascular bundles. 
(c) CAP: there exists segmental consolidation with GGO.}
\end{centering}
\end{figure}

However, CT scans also share several similar visual manifestations between COVID-19 and other types of pneumonia, thus making it difficult and time-consuming for doctors to differentiate among a mass of cases, resulting in about 25$\sim$53\% specificity~\cite{ai2020correlation,fang2020sensitivity}. 
Among them, CAP (community-acquired pneumonia) and influenza pneumonia are the most common types of pneumonia, as shown in Figure~\ref{fig:sample}; therefore, it is essential to differentiate COVID-19 pneumonia from these.  
Recently, Liu \emph{et al.} compared the chest CT characteristics of COVID-19 pneumonia with influenza pneumonia, and found that COVID-19 pneumonia was more likely to have a peripheral distribution, with the absence of nodules and tree-in-bud signs~\cite{liu2020covid}. 
Lobar or segmental consolidation with or without cavitation is common in CAP~\cite{chen2019pulmonary}. 
Although it is easy to identify these typical lesions, the CT features of COVID-19, H1N1 and CAP pneumonia are very diverse.

In the past few decades, artificial intelligence using deep learning (DL) technology has achieved remarkable progress in various computer vision tasks~\cite{krizhevsky2012imagenet,wang2018non,qian2019leader,wang2020fm2u,chen2019hybrid}. 
Recently, the superiority of DL has made it widely favored in medical image analysis. 
Specifically, several studies focus on classifying different diseases, such as autism spectrum disorder ~\cite{heinsfeld2018identification,kong2019classification} or Alzheimer's disease in the brain~\cite{liu2014early,ortiz2016ensembles,jo2019deep}; breast cancers~\cite{albarqouni2016aggnet,bejnordi2017diagnostic,Deng2019}; diabetic retinopathy and Glaucoma in the eye~\cite{sayres2019using,Orlando2020,Fu2020_AGE}; and lung cancer~\cite{sun2016computer,ardila2019end} or pneumonia~\cite{zech2018variable,li2020artificial} in the chest. 
Some efforts have also been made to partition images, from different modalities (\emph{e.g.}, CT, X-ray, MRI) into different meaningful segments~\cite{li2015automatic,guo2019deep,Zhou2020TMI}, including pathology, organs or other biological structures. 

Existing studies~\cite{li2020artificial,wang2020deep,wang2020fully} have demonstrated the promising performance of applying deep learning technology for COVID-19 diagnosis. 
However, as initial studies, several limitations have emerged from these works. 
First of all,~\cite{medseg,fan2020inf,ma2020towards,shi2020review} utilized pixel-wise annotations for segmentation, which require taxing manual labeling. 
This is unrealistic in practice, especially in the event of an infectious disease pandemic. 
Second, performing diagnosis or risk assessment on only slice-level CT images~\cite{shan2020lung,mobiny2020radiologist,hu2020weakly,butt2020deep,jin2020development,wang2020deep,yang2020deep} is of limited value to clinicians.
Since a volumetric CT exam normally includes hundreds of slices, it is still inconvenient for clinicians to go through the predicted result of each slice one by one. 
Although, 3D Convolutional Neural Networks (3D CNNs) are one option for tackling these limitations, their high hardware requirements, computational costs (\emph{e.g.}, GPUs) and training time, make them inflexible for applications~\cite{wu2019deep,butt2020deep,gozes2020rapid}.

To this end, we propose a Multi-task Multi-slice Deep Learning System (M$^3$Lung-Sys) for multi-class lung pneumonia screening, which can jointly diagnose and locate COVID-19 from chest CT images. 
Using the only category labeled information, our system can successfully distinguish COVID-19 from H1N1, CAP and healthy cases, and automatically locate relevant lesions on CT images (\emph{e.g.}, GGO) for better interpretability, which is more important for assisting clinicians in practice. 
To facilitate the above objective, two networks using a 2D CNN are devised in our system. 
The first one is a slice-level classification network, which acts like a radiologist to diagnose from coarse (normal or abnormal) to fine (disease categories) for every single CT slice. 
As the name suggests, it can ignore the temporal information among CT volumes and focus on the spatial information among pixels in each slice. 
Meanwhile, the learned spatial features can be further leveraged to locate the abnormalities without any annotation. 
To recover the temporal information and provide more value to clinicians, we introduce a novel patient-level classification network, using specifically designed refinement and aggregation modules, for diagnosis from CT volumes. 
Taking advantage of the learned spatial features, the patient-level classification network can be trained easily and efficiently. 

In summary, the contributions of this paper are four-fold: 
1) We propose an M$^3$Lung-Sys for multi-class lung pneumonia screening from CT images. 
Specifically, it can distinguish COVID-19 from healthy, H1N1 and CAP cases on either a single CT slice or CT volumes of patients. 
2) In addition to predicting the probability of pneumonia assessment, our M$^3$Lung-Sys is able to simultaneously output the lesion localization maps for each CT slice, which is valuable to clinicians for diagnosis, allowing them to understand why our system gives a particular prediction, rather than simply being fed a statistic. 
3) Compared with 3D CNN based approaches~\cite{9097297,gozes2020rapid}, our proposed system can achieve competitive performance with a cheaper cost and without requiring large-scale training data \footnote{Large scale of patient-level training cases, which are required for 3D CNN based methods, are very difficult to access due to various complex factors, \emph{e.g.}, time limitation and patient privacy. 
However, a small scale of CT volumes can provide plenty of slice-level samples with category labels, which can be utilized in a 2D CNN system.}. 
4) Extensive experiments are conducted on a multi-class pneumonia dataset with 251 healthy people, 245 COVID-19 patients, 105 H1N1 patients and 133 CAP patients. 
We achieve high accuracy of 95.21\% for correctly screening the multi-class pneumonia testing cases. The quantitative and qualitative results demonstrate that our system has great potential to be applied in clinical application.

\begin{table}[!t]
\begin{centering}
\setlength{\tabcolsep}{1.9mm}{
\begin{tabular}{l|c|c|c|c|c|c}
\hline 
\multicolumn{1}{c|}{\multirow{2}{*}{Category}} & \multicolumn{2}{c|}{Patient-level} & \multicolumn{2}{c|}{Slice-level} & \multirow{2}{*}{Age} & Sex\tabularnewline
\cline{2-5}
 & Train  & Test  & Train  & Test & & (M/F) \tabularnewline
\hline 
Healthy  & 149  & 102  & 42,834  & 30,448 & 32.4$\pm$11.8 & 131/120 \tabularnewline
COVID-19 & 149  & 96  & 18,919 & 13,382 & 51.5$\pm$15.9 & 143/102 \tabularnewline
H1N1  & 64  & 41  & 1,098  & 883 & 28.5$\pm$14.6 & 62/43 \tabularnewline
CAP  & 80  & 53  & 7,067  & 4,105 & 48.5$\pm$17.4 & 79/54\tabularnewline
\hline 
\end{tabular}}
\end{centering}
\caption{Summary of training and testing sets.}
\label{Tab:dataset}
\end{table}

\section{DATASET}

\subsection{Patients}
The Ethics Committee of Shanghai Public Health Clinical Center, Fudan University approved the protocol of this study and waived the requirement for patient-informed consent (YJ-2020-S035-01).
A search through the medical records in our hospital information system was conducted, with the final dataset consisting of 245 patients with COVID-19 pneumonia, 105 patients with H1N1 pneumonia, 133 patients with CAP and 251 healthy subjects with non-pneumonia. 
Of the 734 enrolled people with 415 (56.5\%) men, the mean age was 41.8$\pm$15.9 years (range, $2\sim96$ years). The patient demographic statistics are summarized in Table \ref{Tab:dataset}. 
The available CT scans were directly downloaded from the hospital Picture Archiving and Communications Systems (PACS) and non-chest CTs were excluded. 
Consequently, 734 three-dimensional (3D) volumetric chest CT exams are acquired for our algorithm study. 

All the COVID-19 cases (mean age, $51.5\pm15.9$ years; range, $16\sim83$ years) and H1N1 cases (mean age, $28.5\pm14.6$ years; range, $4\sim78$ years) were acquired from January 20 to February 24, 2020 and from May 24, 2009 to January 18, 2010, respectively. 
All patients were diagnosed according to the diagnostic criteria of the National Health Commission of China and confirmed by RT-PCR detection of viral nucleic acids. Patients with normal CT imaging were excluded.

The patients with CAP subjects (mean age, $48.5\pm17.4$ years; range, $8\sim96$ years) and healthy subjects (mean age, $32.4\pm11.8$ years; range, $2\sim73$ years) with non-pneumonia were randomly selected between January 3, 2019 and January 30, 2020. 
All the CAP cases were confirmed positive by bacterial culture, and healthy subjects with non-pneumonia undergoing physical examination had normal CT imaging.

\begin{figure}[!t]
\begin{centering}
\includegraphics[width=0.9\linewidth]{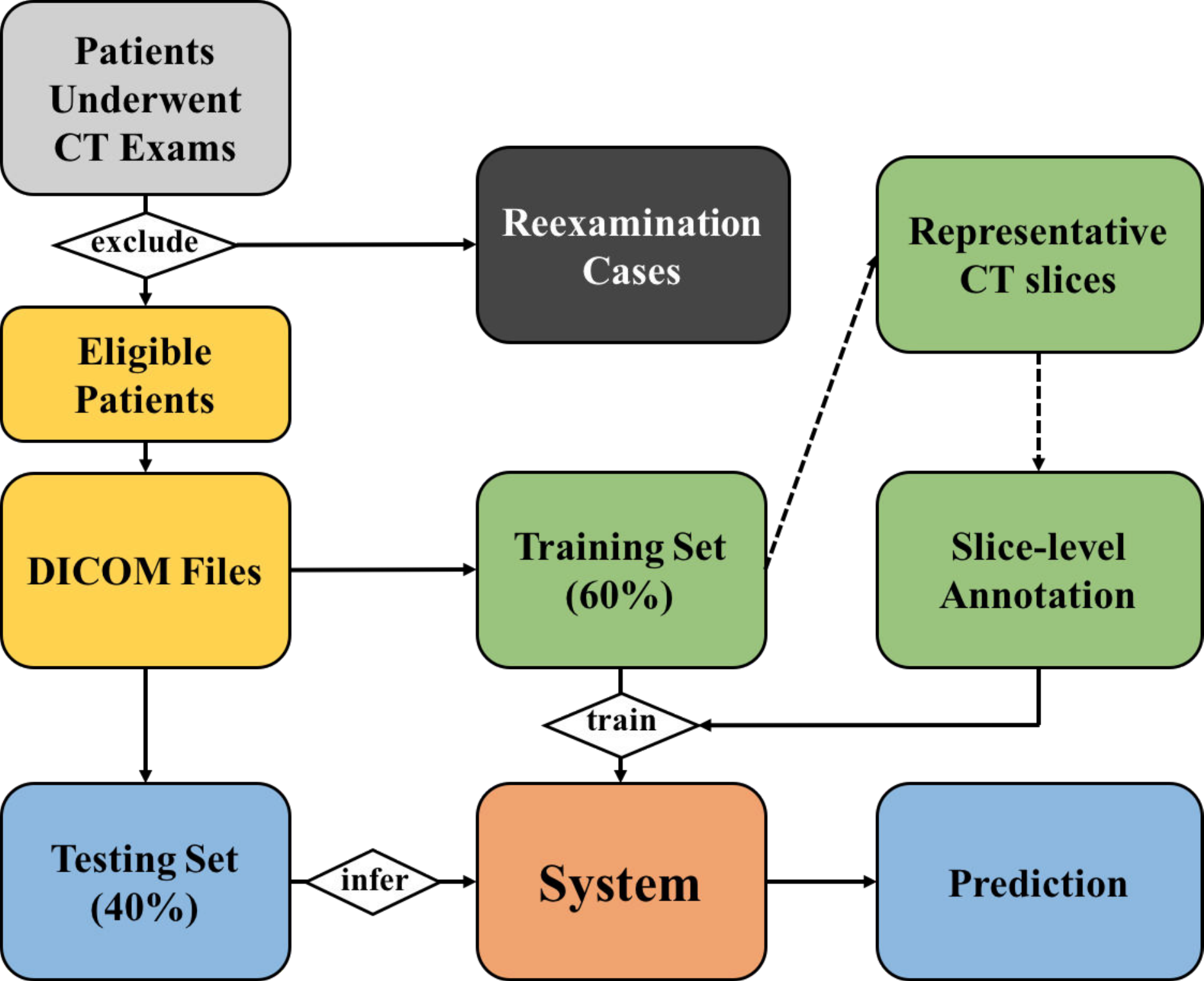} 
\caption{\label{fig:flow_chart} The flow chart of patient selection.}
\end{centering}
\end{figure}

\subsection{Selection and Annotation}
To better improve the algorithm framework and fairly demonstrate the performance, we do not use any CT volumes from re-examination, that is, only one 3D volumetric CT exam per patient is enrolled in our dataset. 
As shown in Figure~\ref{fig:flow_chart}, all eligible patients were then randomized into a training set and testing set, respectively, using random computer-generated numbers. 
Unlike other studies~\cite{gozes2020rapid,yang2020deep} which employ a small number of cases (10\%$\sim$15\%) for testing, we utilize around 40\% of each category to evaluate the effectiveness and practicability of our system. 

The annotation was performed at a patient and slice level. First of all, each CT volume was automatically labeled with a one-hot category vector based on CT reports and clinical diagnosis (\emph{i.e.}, 0: Healthy; 1: COVID-19; 2: H1N1; 3: CAP).
Considering that each volumetric exam contains $512\times512$ images with a varying number of slices from 24$\sim$495, for training,
five experts subsequently annotated each CT slice following four principles:
(1) The quality of annotation is supervised by a senior clinician;
(2) If a slice is determined to have any lesion, label it with the corresponding CT volume's category; 
(3) Except for healthy cases, all slices from other cases considered as normal are discarded;
(4) All slices from healthy people are annotated as Healthy. 
Note that we evaluate our model with the whole CT volume (\emph{i.e.}, realistic and arbitrary-length data), the discarded slices are only removed for training.
Eventually, the number of slices annotated for the four categories is listed in Table~\ref{Tab:dataset}.
The training set was used for algorithm development [n=442; healthy person, n=149; COVID-19 patients, n=149; H1N1 patients, n=64; CAP patients, n=80], and the testing set was used for algorithm testing [n=292; healthy person, n=102; COVID-19 patients, n=96; H1N1 patients, n=41; CAP patients, n=53].

\begin{figure*}
\begin{centering}
\includegraphics[width=0.9\linewidth]{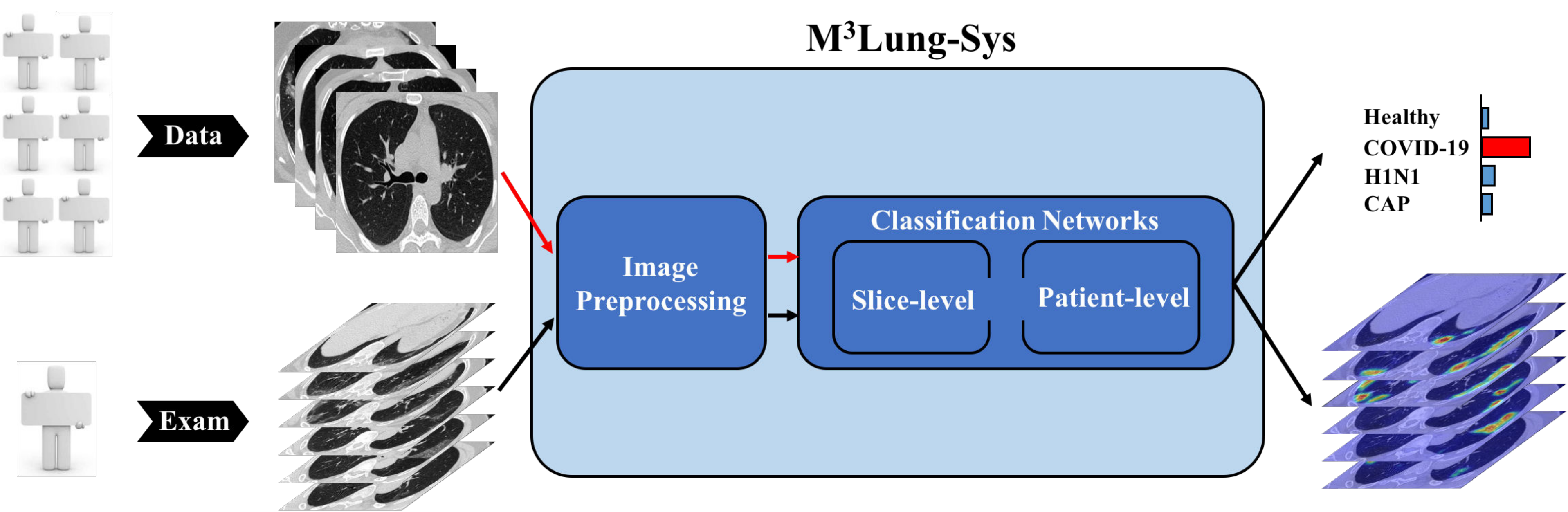} 
\par\end{centering}
\caption{\label{fig:system} The schematic of our M$^3$Lung-Sys.
The red/black arrows indicate the training/inference phases.
Both classification networks are trained separately due to different tasks, but can be used concurrently in an end-to-end manner.
The details of the classification networks are illustrated in Figure~\ref{fig:framework} and Figure~\ref{fig:refinement_aggregation}.}
\end{figure*}

\section{METHODOLOGY}
Figure~\ref{fig:system} shows the schematic of our proposed Multi-task Multi-slice Deep Learning System (M$^{3}$Lung-Sys), which consists of two components, the Image Preprocessing and Classification Network. 
Specifically, the Image Preprocessing receives raw CT exams, and prepare them for model training or inference (in Section~\ref{Image Preprocessing}). 
For the Classification Network, we divide it into two subnets (stages), \emph{i.e.}, slice-level and patient-level classification networks, with the purpose of jointly COVID-19 screening and location. 
Concretely, slice-level classification network is trained with multiple CT slice images and predicts the corresponding slice-level categories (in Section~\ref{Weakly-Supervised Multi-Task Classification Network}), \emph{i.e.}, Healthy, COVID-19, H1N1 or CAP. 
Besides, patient-level classification network only has four layers, and takes a volume of CT slice features, instead of images as input, which are extracted by the former network, to output the patient-level labels (in Section~\ref{Multi-scale Volumetirc Classification Network}). Both classification networks are trained separately due to different tasks, but can be used concurrently in an end-to-end manner for the efficiency.
More importantly, for cases classified as positive (\emph{i.e.}, COVID-19, H1N1 or CAP), our system can locate suspected area of abnormality without any pixel-level annotations (in Section~\ref{Weakly-Supervised Multi-Task Classification Network}).

\subsection{Image Preprocessing}

\label{Image Preprocessing} 
The pixel value of CT images reflects the absorption rate of different human tissues (\emph{e.g.}, bone, lung, kidney) to x-rays, which is measured by Hounsfield Unit (HU)~\cite{brooks1977quantitative}. 
If we directly apply raw images for classification, this will inevitably introduce noise or irrelevant information, such as the characteristics of equipment, making the performance of the model inaccurate and unreliable. Consequently, according to the priors from radiologist, here, we introduce two effective yet straightforward approaches for preprocessing.

\begin{figure}
\begin{centering}
\includegraphics[width=1\linewidth]{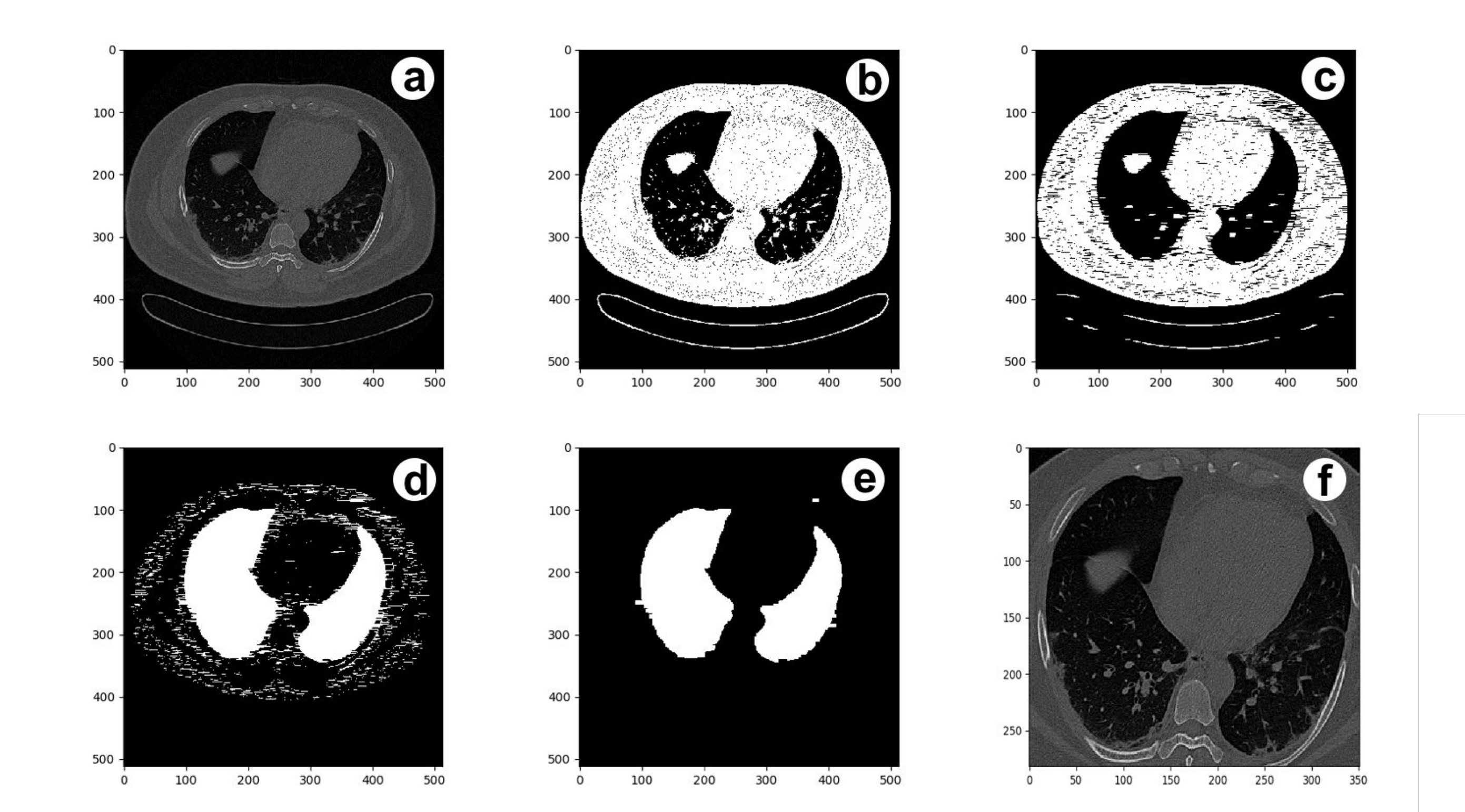}
\par
\caption{\label{fig:segmentation} Detailed procedures of lung cropping during image preprocessing.}
\end{centering}
\end{figure}

\subsubsection{Lung Crop} 
Given chest CT images, lungs are one of the most important organs observed by radiologists to check whether there exist abnormalities. 
Considering the extreme cost and time-consumption of manual labeling, instead of training an extra deep learning network for lung segmentation~\cite{shi2020large,gozes2020rapid,9097297}, we propose a hand-crafted algorithm to automatically subdivide/segment the image into `lungs' and `other', and then crop the area of lungs using the minimum bounding rectangle within a given margin. 
As illustrated in Figure~\ref{fig:segmentation}, the details of the algorithm involved in lung segmentation and cropping are as following: 
\begin{itemize}
\item Step 1: Load the raw CT scan image (Figure~\ref{fig:segmentation} (a)).
\item Step 2: Set a threshold to separate the lung area from others, such as bone and fat (Figure~\ref{fig:segmentation} (b)).  In this paper, we set the threshold of HU as $T_{HU}=-300$.
\item Step 3: To alleviate the effect of `trays', which the patient lays on during CT scanning, we apply a morphological opening operation~\cite{makram2000method} (Figure~\ref{fig:segmentation} (c)). Specifically, we set the kernel size as $1\times 8$.
\item Step 4: Remove the background (\emph{e.g.}, trays, bone and fat) based on 8-Connected Components Labeling~\cite{di1999simple} (Figure~\ref{fig:segmentation} (d)).
\item Step 5: Apply the morphological opening operation again to eliminate the noise caused by Step 3 (Figure~\ref{fig:segmentation} (e)).
\item Step 6: Compute the minimum bounding rectangle with a margin of 10 pixels for lung cropping and then resize the cropped image (Figure~\ref{fig:segmentation} (f)). 
\end{itemize}

\subsubsection{Multi-value Window-leveling}
In order to simulate the process of window-leveling when a radiologist is looking at CT scans, we further apply multi-value window-leveling to all images. 
More concretely, the value of the window center is assigned randomly from $-700$ to $-500$, and the window width is assigned with a constant of $1200$. 
This preprocessing provides at least two benefits: 
(1) generating much more CT samples for training, \emph{i.e.}, data augmentation; 
(2) during inference, the assessment based on multi-value window-leveling CT images will be more accurate and reliable.

\subsection{Slice-level Classification Network}
\label{Weakly-Supervised Multi-Task Classification Network}

After the above-mentioned image preprocessing, slices from CT volumes are first fed into the slice-level classification network. 
Considering the outstanding performance achieved by the residual networks (ResNets)~\cite{he2016deep} on the 1000-category image classification task, we utilize ResNet-50~\cite{he2016deep} as our backbone and initialize it with the ImageNet~\cite{deng2009imagenet} pre-trained weights. 
This network consists of four blocks (a.k.a, ResBlock1$\sim$4) with a total of 50 layers, including convolutional layers and fully connected layers.
Each block has a similar structure, but different number of layers. 
The skip connection and identity mapping functions in the blocks make it more possible to apply deeper layers to learn stronger representations. 
For the purpose of pneumonia classification and alleviating the limitations discussed in Section~\ref{introduction}, we improve the network from three aspects, \emph{i.e.}, multi-task learning for radiologist-like diagnosis, weakly-supervised learning for slice-level lesion localization (attention) and coordinate maps for learning location information, as shown in Figure~\ref{fig:framework}.

\subsubsection{Multi-task Learning}
\label{subsec:multi-task}
Usually, given a CT slice, a radiologist will gradually check for abnormalities and make a decision according to these. 
To act like an experienced radiologist, we introduce a multi-task learning scheme \cite{caruana1997multitask} by dividing the network into two stages.
Specifically, image features obtained from the first three ResBlocks are fed into an extra classifier to determine whether they have any lesion characteristics. 
Then, the features are further passed through ResBlock4 to determine fine-grained category, \emph{i.e.}, Healthy, COVID-19, H1N1 or CAP.

\subsubsection{Weakly-supervised Learning for Lesion Localization}
\label{sec:lesion detection}
\begin{figure}
\begin{centering}
\includegraphics[width=1\linewidth]{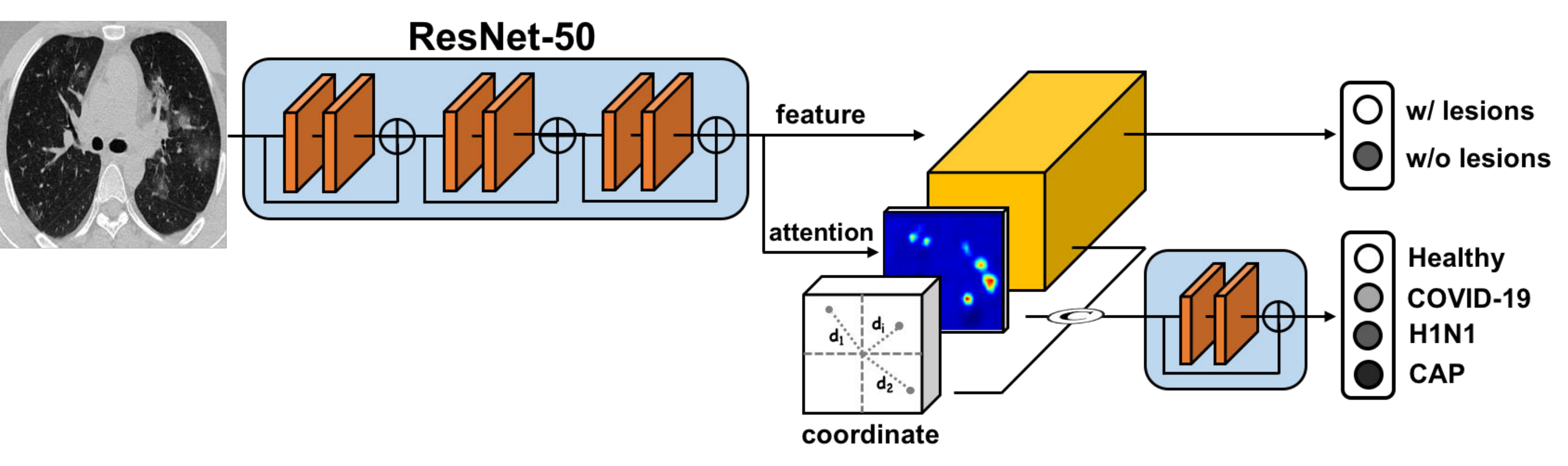} 
\par
\caption{\label{fig:framework} The details of our proposed slice-level classification network. 
We improve the network from three aspects: (1) multi-task learning to diagnose like a radiologist; (2) weakly-supervised learning for slice-level lesion localization (attention); (3) coordinate maps for location information. 
The symbol `\copyright' indicates a concatenation operation.}
\end{centering}
\end{figure}

Instead of using pixel-wise annotations or bounding box labels for learning to locate infection areas, we devise a weakly-supervised learning approach, that is, employ only the category labels.
Specifically, the weights of the extra classifier described in Sec.~\ref{subsec:multi-task} have a dimension of $2\times D$, where $D$ is the dimension of the feature and `2' denotes the number of classes (\emph{i.e.}, `with lesion' and `without lesion'). 
These learned weights can be regarded as two prototypical features of the corresponding two classes. 
Similar to the Class Activation Map~\cite{zhou2016learning}, we first select one prototypical feature according to the predicted class, and then calculate the distance between it and each point of the image feature extracted from the first three ResBlocks.
Intuitively, a closer distance between a point and the prototypical feature of `with lesion' indicates that the area of this point mapping to the input CT slice has a higher probability of being an infection region, \emph{e.g.}, GGO.
As one output of our M$^3$Lung-Sys, such generated location maps are complementary to the final predicted diagnosis and provide interpretability for our network, making the assistance to clinicians more comprehensive and flexible. 
More visualization samples are demonstrated in Figure~\ref{fig:visualization}, \ref{fig:visualization-patient-2} and \ref{fig:visualization-patient-1}. 
Furthermore, we regard the lesion location map as an attention map and take full advantage of it to help the slice-level differential diagnosis, as shown in Figure~\ref{fig:framework}.

\subsubsection{Coordinate Maps} 
From the literature~\cite{bernheim2020chest}, it is known that infections of COVID-19 have several spatial characteristics. 
For example, they frequently distribute bilaterally on lungs, and predominantly in peripheral lower zone. 
Nevertheless, convolutional neural networks primarily extract the features of textures. 
To explicitly capture the spatial information, and inspired by~\cite{liu2018intriguing}, we integrate our slice-level classification network with the coordinate maps ($H \times W \times 3$) containing three channels, where $H$ and $W$ are the height and width of the image feature extracted from the first three ResBlocks, to facilitate the distinction among COVID-19, H1N1 and CAP. 
The first two channels of the coordinate maps are instantiated and filled with the coordinates of $x \in \left[ 0, W \right)$ and $y \in \left[ 0, H \right)$ respectively. And we further apply a normalization to make them fall in the range of $[-1,1]$. The last channel encodes the distance $d$ from the point $(x, y)$ to the center $(0, 0)$, \emph{i.e}, $d=\sqrt{x^{2}+y^{2}}$.
Specifically, these three additional channels are fed into the ResBlock4 together with the image feature and attention map to learn representations with spatial information.

\begin{figure}
\begin{centering}
\includegraphics[width=1\linewidth]{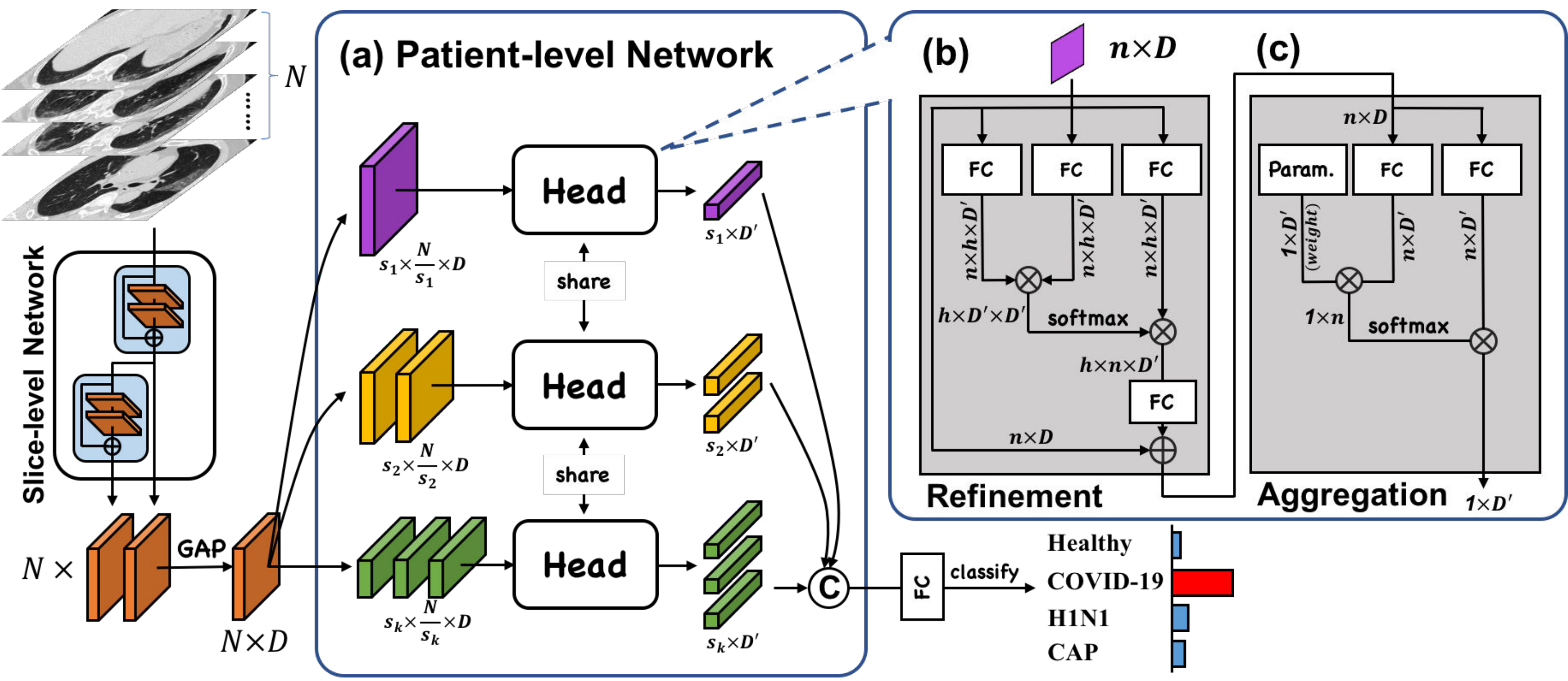} 
\par
\caption{\label{fig:refinement_aggregation} 
The details of our proposed patient-level classification network. It takes a  volume of slice-level features as input, and feeds them into a feature refinement and aggregation head, so that the image features from different slices can be correlated with each other and aggregated into one final feature for patient-level classification.}
\end{centering}
\end{figure}

\subsection{Patient-level Classification Network}
\label{Multi-scale Volumetirc Classification Network}

As mentioned in Section~\ref{introduction}, performing diagnosis or risk assessment on only slice-level CT images is of limited value to clinicians. 
Although several studies~\cite{9097297,gozes2020rapid} have been proposed to take advantage of temporal information with 3D CNNs for patient-level differential diagnosis, they require thousands of patient-level data for deep model training, which makes the cost particularly high.
To overcome these limitations, we further propose a patient-level classification network. 
It takes a volume of CT slice-level features as input rather than 3D images, and only comprises four layers, allowing it to be trained with lower hardware, time and data cost. 
Details will be described below. 
Note that we concatenate the features from ResBlock3 and ResBlock4 in Section~\ref{Weakly-Supervised Multi-Task Classification Network} as the input.

\subsubsection{Feature Refinement and Aggregation Head.}
Inspired by~\cite{vaswani2017attention,wang2018non}, we introduce a three-layer head to conduct feature refinement and aggregation, so that the image features from different slices can be correlated with each other and aggregated into one final feature for patient-level classification. The key intuition behind this is to utilize the attention mechanism to exploit the correlation between different CT slices, and accomplish the refinement and aggregation based on the explored correlation. As shown in Figure~\ref{fig:refinement_aggregation}, the head includes a feature refinement module with two layers and a feature aggregation module with one layer, the structures of which are similar.

Formally, for the feature refinement module, given a volume of CT image features with the feature dimension of $D$, we first utilize three parallel FC layers to map the input to three different feature spaces for dimension reduction ($D^{'} < D$) and self-attention~\cite{wang2018non}. Then, we calculate the distance, as attention, between each pixel in different slices using features from the first two spaces, and refine the features from the last space based on the attention. 
Finally, another FC layer is employed to expand the feature dimension back to $D$, so that the skip-connection operation~\cite{he2016deep} can be applied.
Similar to~\cite{vaswani2017attention}, we also apply a multi-head mechanism \footnote{The multi-head mechanism refers to the dimension of `h' in Eq.~\ref{eq: mapping} and~\ref{eq: attention}.} to strengthen the refinement ability. 
Without loss of generality, we define the input volume feature as $\mathbf{F}\in \mathbb{R}^{n\times D}$, where $n$ is the number of slices, and the overall formulation of our refinement module can be expressed as,

\begin{align}
& \mathbf{F}_{1}=f_{\theta_{1}}\left(\mathbf{F}\right),~~
\mathbf{F}_{2}=f_{\theta_{2}}\left(\mathbf{F}\right),~~
\mathbf{F}_{3}=f_{\theta_{3}}\left(\mathbf{F}\right), \label{eq: mapping} \\
& \mathbf{H} = \frac{ \mathbf{F}^{\mathrm{T}}_{1}\mathbf{F}_{2} }{ \sum_{j=1}^{D^{'}}\mathbf{F}^{\mathrm{T}}_{1}\mathbf{F}_{2}[:,j]}, \; with \; \mathbf{F}^{\mathrm{T}}_{1}\mathbf{F}_{2} \in \mathbb{R}^{h \times D^{'}\times D^{'}} \label{eq: attention} \\
& \mathbf{F}_{out} = f_{\theta_{4}}\left( \mathbf{F}_{3}\mathbf{H} \right),
\; with \; \mathbf{F}_{out} \in \mathbb{R}^{n \times D}
\label{eq: fc}
\end{align}


\noindent where $\mathbf{H}$ indicates the correlation between each pixel in different slices, $f_{\theta_{i}}$ indicates the $i$-th FC layer with parameter $\theta_{i}$, and we omit the reshape operation for simplicity.

With regard to the feature aggregation module, its structure, as well as the equations, is similar to the refinement module, except that we remove the multi-head mechanism and the last FC layer, and replace the first FC layer with a learnable parameter $k\in \mathbb{R}^{1\times D^{'}}$, so that the total number of $N$ CT image features can be aggregated into one. Details can be found in Figure~\ref{fig:refinement_aggregation} (c).

\subsubsection{Multi-scale Learning} 
If the number of slices with lesions in the early stage is relatively small (\emph{i.e.}, $<N $), this may result in key information being leaked when performing feature aggregation from $N\times D$ to $1\times D$. Therefore, we introduce a multi-scale learning mechanism to aggregate features from different scales. 
As illustrated in Figure~\ref{fig:refinement_aggregation} (a), given a set of scales $S=[s_{1}, s_{2}, \dots, s_{k}]$, for each scale $s_{j}$, we first divide the input feature $\mathbf{F}\in \mathbb{R}^{N\times D}$ evenly into $s_{j}$ parts, $\{\mathbf{F}_{i} \in \mathbb{R}^{\frac{N}{s_{j}} \times D}~\vert~ i\in[1,2,\dots,s_{j}] \}$.
Then, a shared feature refinement and aggregation head is applied to each part. In the end, we concatenate a set of aggregated features from all parts of different scales, and feed it into one FC layer to reduce the dimension from $\sum_{j=1}^{k}s_{k}D $ to $D$ as the final patient-level feature for classification.

\begin{table*}
\begin{centering}
\setlength{\tabcolsep}{1mm}{ %
\begin{tabular}{l|ccc|ccc|ccc|ccc|cccc}
\hline 
\multicolumn{1}{c|}{\multirow{2}{*}{Metrics}} & \multicolumn{3}{c|}{Healthy} & \multicolumn{3}{c|}{COVID-19}  & \multicolumn{3}{c|}{H1N1}  & \multicolumn{3}{c|}{CAP} &
\multicolumn{4}{c}{Overall}\tabularnewline
\cline{2-17}
& Sen. & Spec. & AUC & Sen. & Spec. & AUC & Sen. & Spec. & AUC & Sen. & Spec. & AUC & Acc. & FPE & FNE & FDPE \tabularnewline
\hline 
COVNET & 51.92 & 96.88 & 96.30 & 95.88 & 80.74 & 95.15 & 58.33 & 94.86 & 84.82 & 60.71 & 85.83 & 77.89 & 68.84 [63.70$\sim$74.32] ($p<0.001$) & 48.08 & 3.13 & 18.97 \tabularnewline
\hline 
Med3D-50 & 69.70 & 90.59 & 91.44 & 75.00 & 88.83 & 88.22 & 93.02 & 97.24 & 97.70 & 79.39 & 90.96 & 92.09 & 76.37 [71.57$\sim$80.83] ($p<0.001$) & 30.30 & 9.41 & 10.45 \tabularnewline
Med3D-18 & 93.26 & \textbf{100} & 99.99 & 95.05 & \textbf{99.49} & 99.81 & 92.86 & \textbf{100} & \textbf{100} & \textbf{98.16} & 93.67 & \textbf{99.41} & 94.52 [91.78$\sim$96.92] ($p<0.001$) & 6.74 & \textbf{0.00} & 4.62 \tabularnewline
DeCovNet  & \textbf{99.03} & \textbf{100} & \textbf{100} & 95.96 & 96.04 & 98.87 & \textbf{100} & 98.03 & 99.85 & 75.51 & 97.93 & 96.51 & 93.83 [91.10$\sim$96.58] ($p<0.001$) & \textbf{0.96} & \textbf{0.00} & 8.91 \tabularnewline

\hline
Ours  & 97.17 & 95.86 & 98.88 & \textbf{98.99} & 97.49 & \textbf{99.93} & \textbf{100} & 99.61 & \textbf{100} & 81.19 & \textbf{100} & 97.71 & \textbf{95.21} [92.81$\sim$97.26] & 2.83 & 4.15 & \textbf{1.57} \tabularnewline
\hline
\end{tabular}} 
\par\end{centering}
\caption{\label{Tab:Results on patient-level} Comparing our model with several competitors on patient-level diagnosis. `FPE', `FNE' and `FDPE' denote the metrics of false positive error, false negative error and false disease prediction error, respectively. The numbers in square brackets represent the 95\% confidence interval. `$p$' means the $p$-value compared our model with other competitors.}
\end{table*}

\section{EXPERIMENTS}

\begin{figure*}
\begin{centering}
\includegraphics[width=1\linewidth]{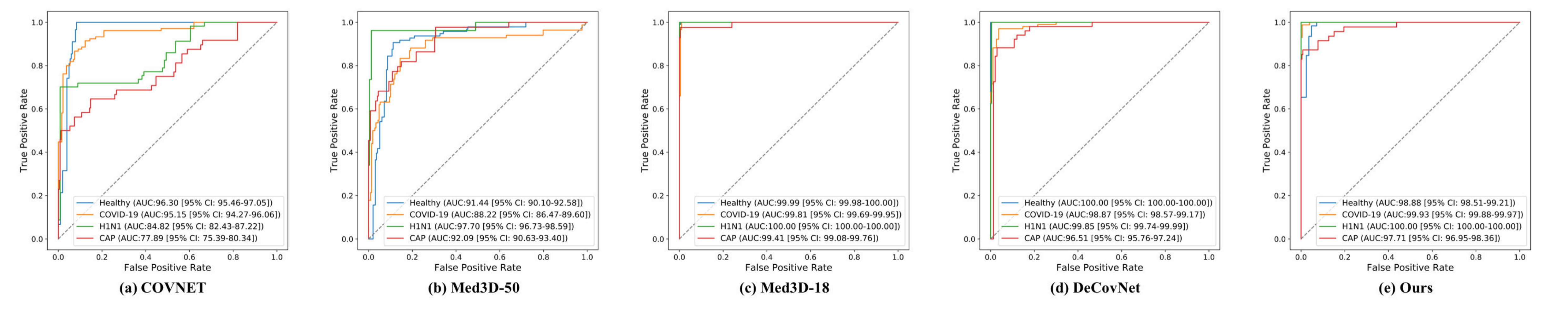}
\caption{\label{fig:roc_patient} ROC plots for patient-level classification. Best viewed in color and zoom in.}
\end{centering}
\end{figure*}

\subsection{Implementation Details}

We implement our framework with PyTorch~\cite{paszke2017automatic}. All CT slices are resized to $512\times512$. We set the hyper parameters of $D^{'}$ and $h$ as 512 and 12 respectively, and use four scales in the feature refinement and aggregation head, \emph{i.e.}, $S=[1,2,3,4]$. For training, the slice-level/patient-level classification network is trained with two/one NVIDIA 1080Ti GPUs for a total of $110$/$90$ epochs, the initial learning rate is $0.01$/$0.001$ and gradually reduces by a factor of $0.1$ every 40/30 epochs. 
Both classification networks are trained separately with the standard cross-entropy loss function. Random flipping is adopted as data argumentation. 
During inference, our system is an end-to-end framework since the input of the patient-level classification network is the output of the slice-level one, so that it can be applied effectively. We additionally set the window center as $[-700, -600, -500]$ for multi-scale window-leveling and average the final predicted features/scores for assessment.

\subsection{Statistical Analysis}

For the statistical analysis, we apply lots of metrics to thoroughly evaluate the performance of the model, following standard protocol.
Concretely, `sensitivity', known as true positive rate (TPR), indicates the percentage of positive patients with correct discrimination. 
Referred as true negative rate (TNR), `specificity' represents the percentage of negative persons who are correctly classified. 
`accuracy' is the percentage of the number of true positive (TP) and true negative (TN) subjects.
`false positive/negative error' (FPE/FNE) measures the percentage of negative/positive persons who are misclassified as positive/negative. 
`false disease prediction error' (FDPE) calculates the percentage of positive persons whose disease types (\emph{i.e.}, COVID-19, H1N1 or CAP) are predicted incorrectly.
Receiver operating characteristic curves (ROC) and area under curves (AUC) are used to show the performance of classifier.
We also report the $p$-values compared our model with other competitors to demonstrate the significance level \footnote{In this paper, we utilize the method of bootstrap to sample $m$ groups of test sets with replacement ($m$ is very large, $1000$, for example), and then calculate the $p$-values from $2 \times m$ groups of results, which are evaluated by our model and other competitors, respectively.}.

\subsection{Experimental Results}
\subsubsection{Patient-level Performance}

The main purpose of our system is to assist the diagnosis of COVID-19 at a patient level rather than slice level~\cite{shan2020lung,mobiny2020radiologist,hu2020weakly,butt2020deep,jin2020development}, which is more significant and practical in the real-world applications. 
Therefore, we first evaluate our system on the patient-level testing set with 102 Healthy, 96 COVID-19, 41 H1N1 and 53 CAP cases. 
The competitors include one 2D CNN based method of COVNet \cite{li2020artificial} and three 3D CNN based models as Med3D-50 \cite{chen2019med3d}, Med3D-18 \cite{chen2019med3d} and DeCovNet \cite{wang2020weakly}. 
The results are shown in Table~\ref{Tab:Results on patient-level} and Figure~\ref{fig:roc_patient}. 

For the overall performance, our system achieves 95.21\% on accuracy, with only 2.83\% and 4.15\% in false positive error and false negative error, respectively. 
Although it may be difficult for clinicians to differentiate COVID-19 from other kinds of viral pneumonia or CAP pneumonia according to CT features, our system, as expected, only gets confused on a small number of cases, \emph{i.e.}, 1.57\% in false disease prediction error, which beats the second best model by a margin of 3.05\%. 
Similarly, for H1N1, it obtains approximately 99.6\% in both sensitivity and specificity, which is definitely a promising performance. 
Moreover, our system significantly improves the sensitive of COVID-19 from 95\% to 99\% comparing with Med3D-18 and DeCovNet. 
However, we observe that the sensitivity or specificity of Healthy is relatively inferior to Med3D-18 and DeCovNet by approximately 2$\sim$4 points, it seems our model is a little oversensitive to noise. 
On the other hand, Med3D-50 achieves much worse performance at most of metrics unexpectedly, especially in sharp contrast to Med3D-18. 
Our explanation is that it may be difficult to train a 3D CNN with such large parameters and limited dataset, which is consist with our motivation of using 2D CNN based network.

\begin{table}
\begin{centering}
\setlength{\tabcolsep}{1.5mm}{ %
\begin{tabular}{l|c|c|c|c|c}
\hline 
\multicolumn{1}{c|}{\multirow{1}{*}{Methods}} & Epoch & Input Size & GPUs & Time(h) & Data \tabularnewline
\hline
COVNet & 110 & 4$\times$3$\times$64$\times$256$\times$256 & 4 & $\approx$2.5 & 442\tabularnewline
\hline 
Med3D-50 & 220 & 6$\times$1$\times$128$\times$256$\times$256 & 6 & $\approx$4.5 & 442\tabularnewline
Med3D-18 & 220 & 8$\times$1$\times$128$\times$256$\times$256 & 4 & $\approx$4 & 442\tabularnewline
DeCovNet & 220 & 16$\times$1$\times$128$\times$256$\times$256 & 4 & $\approx$2.5 & 442\tabularnewline
\hline
Slice-level & 110 & 32$\times$3$\times$1$\times$512$\times$512 & 2 & $\approx$2 & 70k\tabularnewline
Patient-level & 90 & 16$\times$3072$\times$N$\times$1$\times$1 & 1 & $\approx$0.3 & 442\tabularnewline
\hline 
\end{tabular}}
\par\end{centering}
\caption{\label{Tab:Comparsion_Cost} The comparison of computational costs, \emph{i.e.}, the max training epoch, the input size, the required GPUs, training time and the available training data, between our system and several competitors.
The shape of input size is $B\times C\times T\times H \times W$, where $B$, $C$, $T$, $H$, $W$ mean the number of batch-size, channel, slice, image height and width, respectively. 
The `N' denotes the arbitrary number of slice when training patient-level classification network and the unit of the `Data' is the slice/volume for `Slice-level'/other methods.
Note that the input size is limited to the GPU memory. 
}
\end{table}

\begin{figure}[!t]
\begin{centering}
\includegraphics[width=0.9\linewidth]{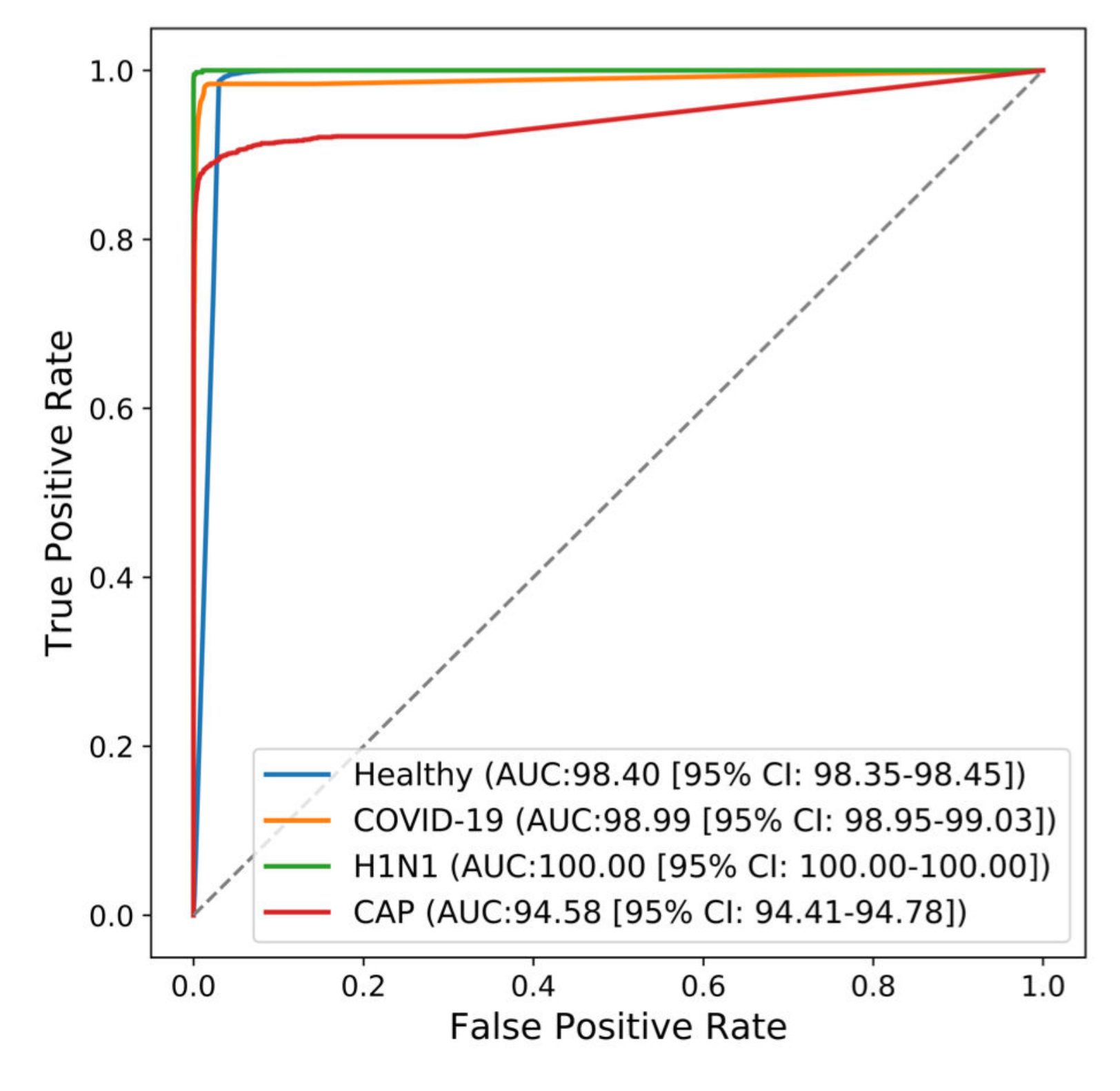}
\caption{\label{fig:roc_slice} ROC plots for slice-level classification.}
\end{centering}
\end{figure}

\begin{table*}
\begin{centering}
\setlength{\tabcolsep}{4mm}{ %
\begin{tabular}{l|c|c|c|c}
\hline 
\multicolumn{1}{c|}{\multirow{1}{*}{Methods}} & Accuracy & FPE & FNE & FDPE\tabularnewline
\hline 
Non-parametric Assessment & 94.18 [91.44$\sim$97.58] ($p<0.001$) & 10.66 & 0.0 & 3.11 \tabularnewline
\hline
Max pooling & 92.12 [89.04$\sim$95.21] ($p<0.001$) & 2.91 & 6.15 & 4.12\tabularnewline
Single-scale + A. Head & 93.15 [90.06$\sim$95.56] ($p<0.001$) & 4.95 & 5.32 & 2.59 \tabularnewline
Multi-scale + A. Head & 94.18 [91.44$\sim$96.58] ($p<0.001$) & 2.89 & 5.22 & 2.04 \tabularnewline
\hline
Multi-scale + R.\&A. Head (Ours) & \textbf{95.21} [92.81$\sim$97.26] & \textbf{2.83} & \textbf{4.15} & \textbf{1.57}\tabularnewline
\hline 
\end{tabular}}
\par\end{centering}
\caption{\label{Tab:Ablation-Patient} Improvements of different components in the patient-level classification network. 
`R.\&A.~Head' is our proposed refinement and aggregation head, and `A.~Head' is a variant without the refinement module. 
`Non-parametric Assessment' denotes that we perform patient-level classification only with several non-parametric mathematical operations. `FPE', `FNE' and `FDPE' indicate the metric of false positive error, false negative error and false disease prediction error, respectively. 
The numbers in square brackets represent the 95\% confidence interval.
`$p$' means the $p$-value compared our model with other competitors.}
\end{table*}

In addition to performance, we also compare the computation cost between our system and other competitors. 
As shown in Table~\ref{Tab:Comparsion_Cost}, our M$^3$Lung-Sys takes full advantage of training data (CT slices and volumes) and has the lowest computation cost, including training time and GPU requirement. Combining with the results in Table~\ref{Tab:Results on patient-level} and Table~\ref{Tab:Comparsion_Cost}, our method can achieve better performance with less computing resources, which is more practical for assisting diagnosis.

\subsubsection{Slice-level Performance}
Another advantage of our proposed M$^3$Lung-Sys is that we can flexibly switch whether the input is CT slices or volumes, \emph{i.e.}, slice-level or patient-level diagnosis. 
Naturally, we further evaluate our model on slices, using the total of 48,818 CT slices from the four categories (\emph{i.e.}, Healthy, COVID-19, H1N1 and CAP) for testing. 
As shown in Figure~\ref{fig:roc_slice}, our model achieves 98.40\%, 98.99\%, 100.00\% and 94.58\% in AUC for the four categories, respectively. 
This strongly demonstrates the superiority of our proposed M$^3$Lung-Sys on slice-level diagnosis.

\begin{table*}
\begin{centering}
\setlength{\tabcolsep}{2mm}{
\begin{tabular}{l|ccc|c|c|c|c}
\hline 
\multicolumn{1}{c|}{Method} & Multi-task & Attention & Coordinate & Accuracy  & FPE  & FNE  & FDPE\tabularnewline
\hline 
ResNet-50 &  &  &  & 89.04 [85.27$\sim$92.47] ($p<0.001$) & 22.46 & 1.04 & 3.65 \tabularnewline
\hline
& \ding{52} & & & 92.12 [89.04$\sim$94.86] ($p<0.001$) & 13.40 & 0.00 & 4.74 \tabularnewline
& \ding{52} & \ding{52} & & 93.49 [90.41$\sim$95.89] ($p<0.001$) & 11.88 & 0.00 & 3.57 \tabularnewline
\hline
Ours  & \ding{52} & \ding{52} & \ding{52} & \textbf{94.18} [91.44$\sim$97.58] & \textbf{10.66} & \textbf{0.00} & \textbf{3.11} \tabularnewline
\hline 
\end{tabular}}
\par\end{centering}
\caption{\label{Tab:ablation-patient-2} Improvements of different components in the slice-level classification network.
`FPE', `FNE' and `FDPE' indicate the metrics of false positive error, false negative error and false disease prediction error, respectively. 
The numbers in square brackets represent the 95\% confidence interval.
`$p$' means the $p$-value compared our model with other competitors.}
\end{table*}

\subsection{Ablation Study}
\subsubsection{Improvements in Patient-level Classification Network}
It is worth mentioning that our proposed slice-level classification network is strong and the extracted features are very discriminative. 
Even without parameters, simple mathematic operations can obtain competitive results on patient-level diagnosis.
Meanwhile, the proposed multi-scale mechanism and refinement and aggregation head are able to further boost performance.
To verify this, as shown in Table~\ref{Tab:Ablation-Patient}, we conduct experiments to demonstrate improvements with different variants of the patient-level classification network. 
More specifically, `Non-parametric Assessment' denotes a simple variant without parameters for differential diagnosis (we refer readers to \ref{app:Non-parametric Assessment} for details). 
`Max pooling' indicates that the input features are directly aggregated by a max pooling operation. 
`Single-scale + A. Head' refers to a variant without the multi-scale mechanism (\emph{i.e.}, S=[3]) and feature refinement module. 
`Multi-scale + A. Head' is similar to the previous model but applies the multi-scale strategy (\emph{i.e.}, S=[1,2,3,4]). 

\begin{figure}
\begin{centering}
\includegraphics[width=1\linewidth]{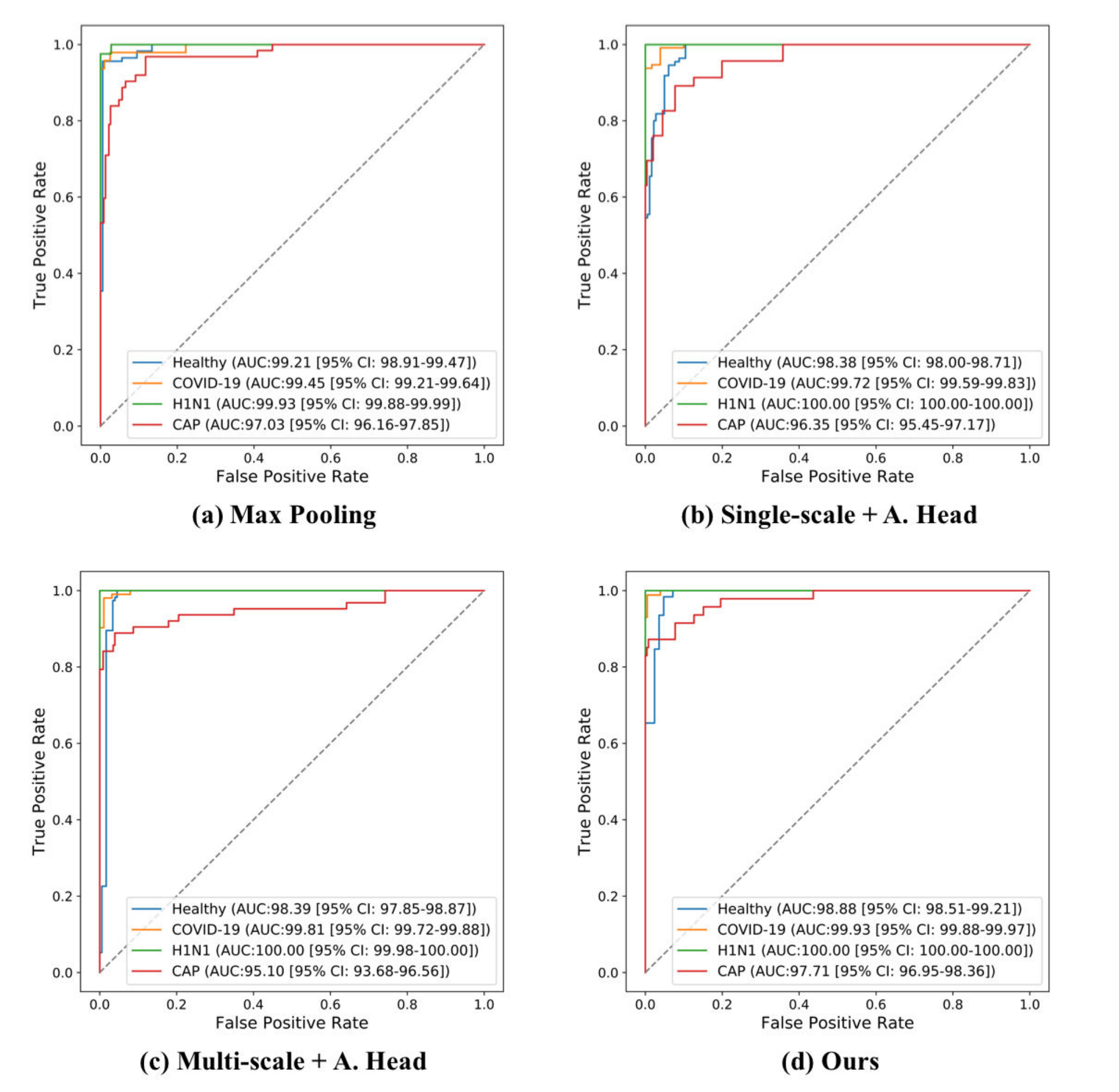} 
\caption{\label{fig:roc_p_ablation} The ROC curves of different variants in patient-level classification network. 
`R.\&A.~Head' is our refinement and aggregation head, and `A.~Head' is a variant without the refinement module. 
Best viewed in color and zoom in.}
\end{centering}
\end{figure}

From the results in Table~\ref{Tab:Ablation-Patient} and Figure~\ref{fig:roc_p_ablation}, we highlight the following observations:
(1) Using only the non-parametric assessment method, we can achieve competitive results of 94.18\% in accuracy, which suggests the stronger feature representations acquired by our slice-level classification network. 
However, a big performance gap between `false positive error' and `false negative error' also reflects its inferior robustness, since a higher value of hyper-parameter $T$ may result in more healthy cases being misdiagnosed due to some noise.
(2) From the results in the second row to the last, the performance on all metrics improves gradually with more and more specifically designed components, which clearly demonstrates the benefits of our proposed feature refinement and aggregation head and multi-scale mechanism. 
(3) We notice that the false positive error gets worse when applying the method of `Single-scale + A. Head', and it decreases dramatically when involving multi-scale mechanism.
We argue that this does not suggest the inferiority of our proposed `A. Head' (since the overall accuracy is improved by 1\%), but reflects the importance and rationality of the multi-scale mechanism.

\subsubsection{Improvements in Slice-level Classification Network}
To explicitly demonstrate the advantages of our improvements in slice-level classification network, we compare it with several competitors on patient-level diagnosis. Without loss of generality, we choose the method of non-parametric assessment with $T=0.99$ as the  patient-level classification network.
Concretely, since the backbone of our slice-level classification network is ResNet-50, which is widely adopted by other works~\cite{li2020artificial,hu2020weakly},
we directly train a vanilla ResNet-50 for four-way classification as a baseline.
Based on this, we conduct further experiments by gradually adding different improvements, including multi-task learning, lesion location (attention) maps and coordinate maps. 

All results are listed in Table~\ref{Tab:ablation-patient-2}.
Compared with the baseline, our model achieves significant improvement in all four metrics. 
For example, the overall accuracy is improved from 89.04\% to 94.18\% and the proportion of false positive is reduced effectively by 12 points. 
Although we obtain a few more failure cases on false disease predication when utilizing the multi-task mechanism, the number of both false positive and false negative samples is dramatically reduced. 
The two-task approach acting like a radiologist is expected to better distinguish between healthy people and patients. Furthermore, if we introduce the coordinate maps, fewer positive samples are misclassified as the wrong type of disease and some negative cases with noise are correctly diagnosed as positive, resulting in a decrease in both false positive and false disease prediction error. 
These results clearly suggest that the components in slice-level classification network play important roles in extracting discriminative features, \emph{e.g.}, attention maps for awareness of small lesions, and coordinate maps for capturing location diversity among different types of pneumonia.

\subsection{Visualizations of Lesion Localization}
\label{sec:Lesion Detection}
\begin{figure}
\begin{centering}
\includegraphics[width=1\linewidth]{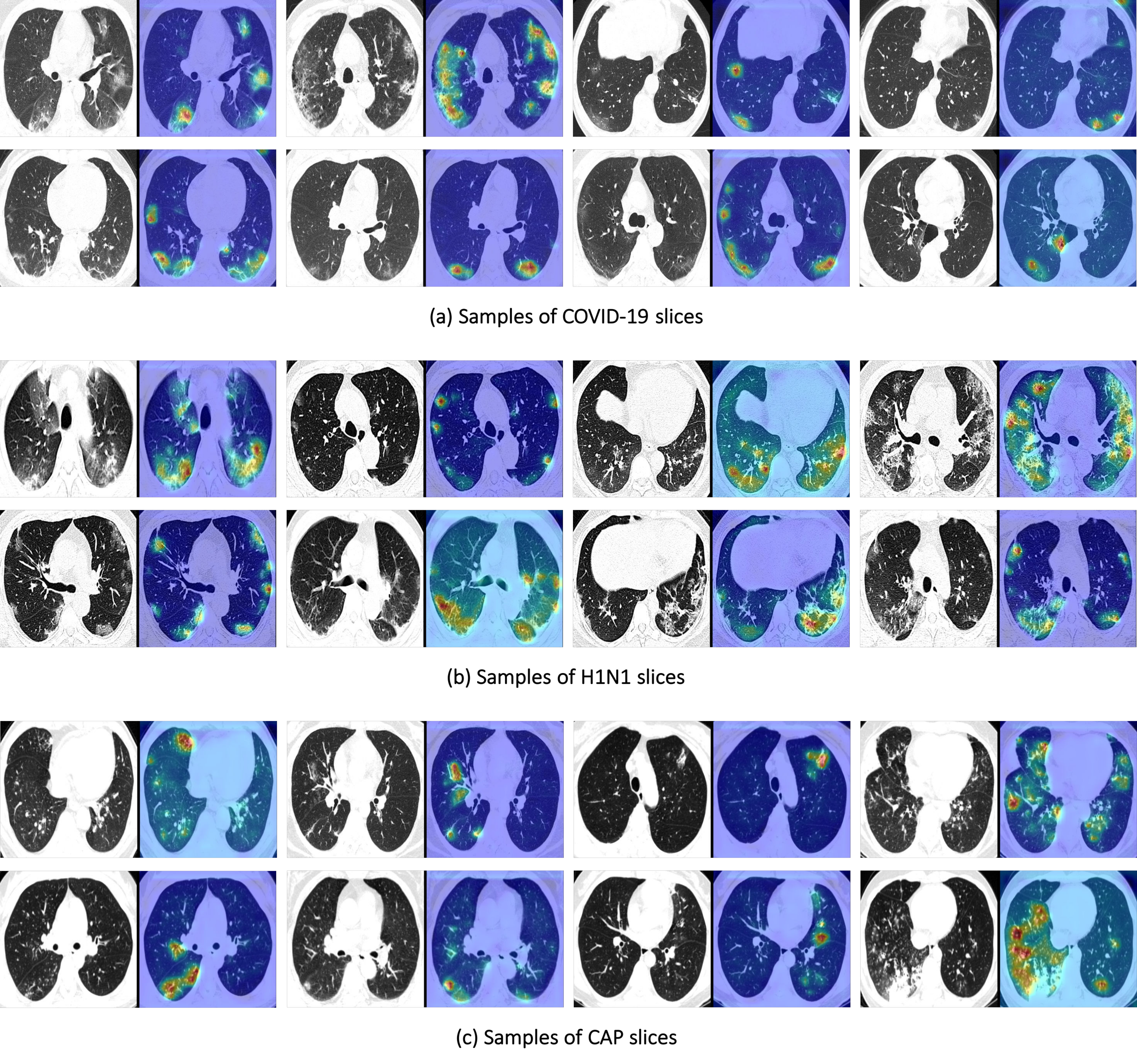} 
\par
\caption{\label{fig:visualization} The visualizations of location maps from COVID-19, H1N1 and CAP cases. Each row has four groups. 
In each group, the left image is the raw CT slice and the right one shows the abnormality areas. Best viewed in color and zoom in.}
\end{centering}
\end{figure}

As one of its contributions, our M$^3$Lung-Sys can implement lesion localization using only category labels, \emph{i.e.}, weakly-supervised learning. 
To qualitatively evaluate this, we randomly select several CT slices of three categories from the testing set, and show the visualizations of lesion location maps in Figure~\ref{fig:visualization}. 
For each group, the left image is the raw CT slice after lung cropping, and the right image depicts the detected area of abnormality. 
Note that a warmer color indicates that the corresponding region has a higher probability of being infected. 
Several observations can be made from Figure~\ref{fig:visualization}:
\begin{enumerate}
\item First of all, the quality of location maps are competitive. 
All highlighted areas are concentrated on the left or right lung region, and most abnormal manifestations, such as ground-glass capacities (GGO), are completely captured by our model, which is trained without any pixel-level or bounding box labels. 
In addition, there is no eccentric area being mistaken as a lesion, such as vertebra, skin or other tissues. 
Our system can even precisely detect small lesions with relatively high response, as shown in the top-right image of Figure~\ref{fig:visualization} (a).
Above all, our system can achieve visual localization of abnormal areas with good interpretability, which is crucial for assisting clinicians in diagnosis and improving the efficiency of medical systems.

\item Second, we found that the location map results are consist with the experience or conclusions of radiologists.
Several studies have found that COVID-19 typically presents GGO with or without consolidation in a predominantly peripheral distribution~\cite{shi2020radiological,song2020emerging,tang2020comparison}, which has already been used as guidance for COVID-19 diagnosis endorsed by the Society of Thoracic Radiology, the American College of Radiology, and RSNA~\cite{simpson2020radiological}. 
In contrast, H1N1 pneumonia most commonly presents a predominantly peribronchovascular distribution~\cite{kang2012computed,li2011pneumonia}. 
Lobar or segmented consolidation and cavitation suggest a bacterial etiology~\cite{chen2019pulmonary}. 
Therefore, the visualizations of location maps can reflect the characteristics of lesion distributions in some ways, which may be a valuable indicator for clinicians in analyzing or differentiating these three diseases. 
\end{enumerate}

\begin{figure}
\begin{centering}
\includegraphics[width=1\linewidth]{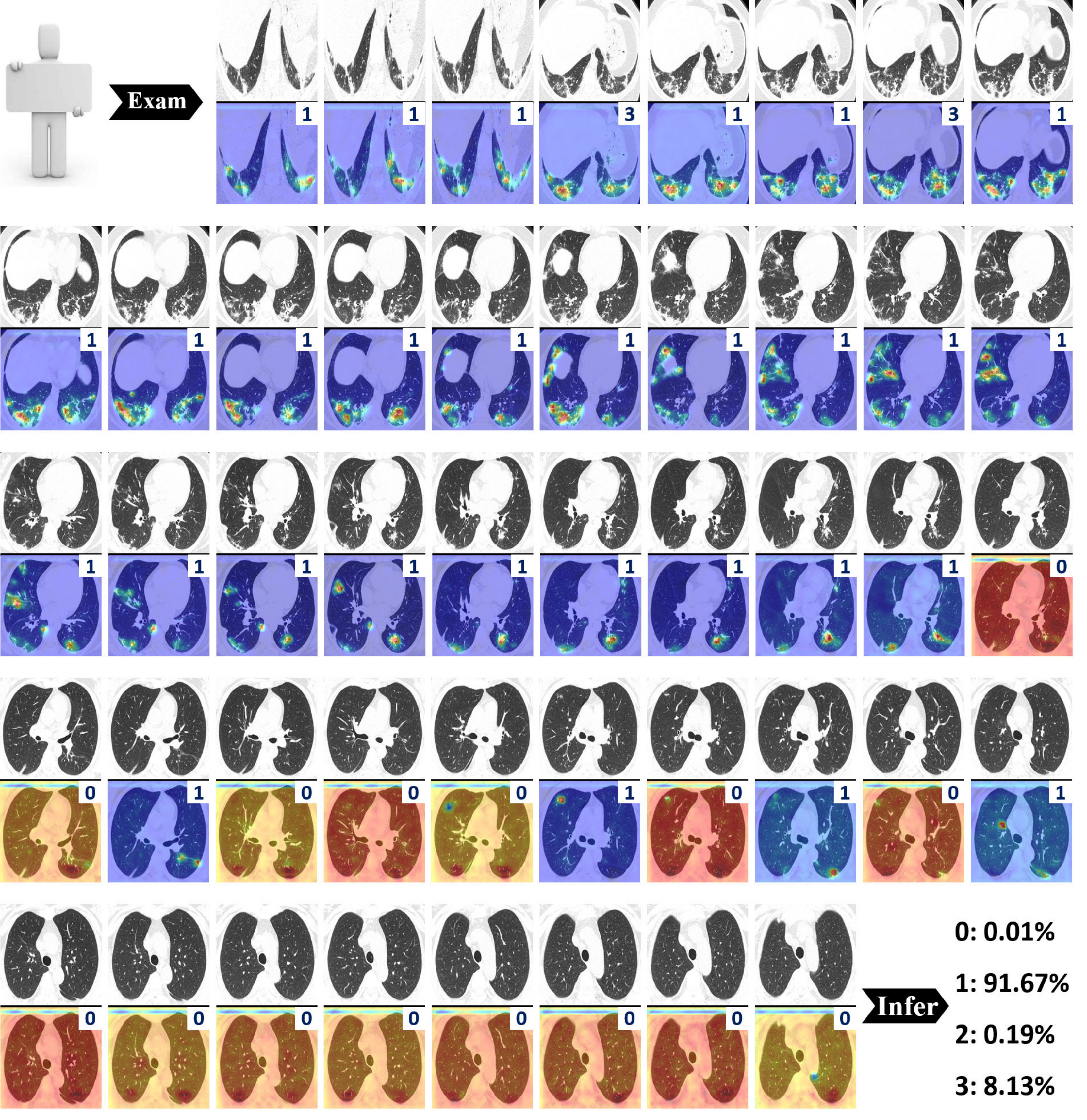} 
\par\end{centering}
\caption{\label{fig:visualization-patient-2} The visualizations of our system outputs for one COVID-19 patient. 
We show a CT sequence by sampling every five slices. 
The lesion location maps, with the predicted slice-level diagnosis on the upper right, are attached at the bottom of the corresponding raw CT slices. 
At the end of the sequence, the probability of each category, predicted by the patient-level classification network is provided, \emph{i.e.}, 0: Healthy; 1: COVID-19; 2: H1N1; 3: CAP. Best viewed in color and zoom in.}
\end{figure}

\begin{figure}
\begin{centering}
\includegraphics[width=1\linewidth]{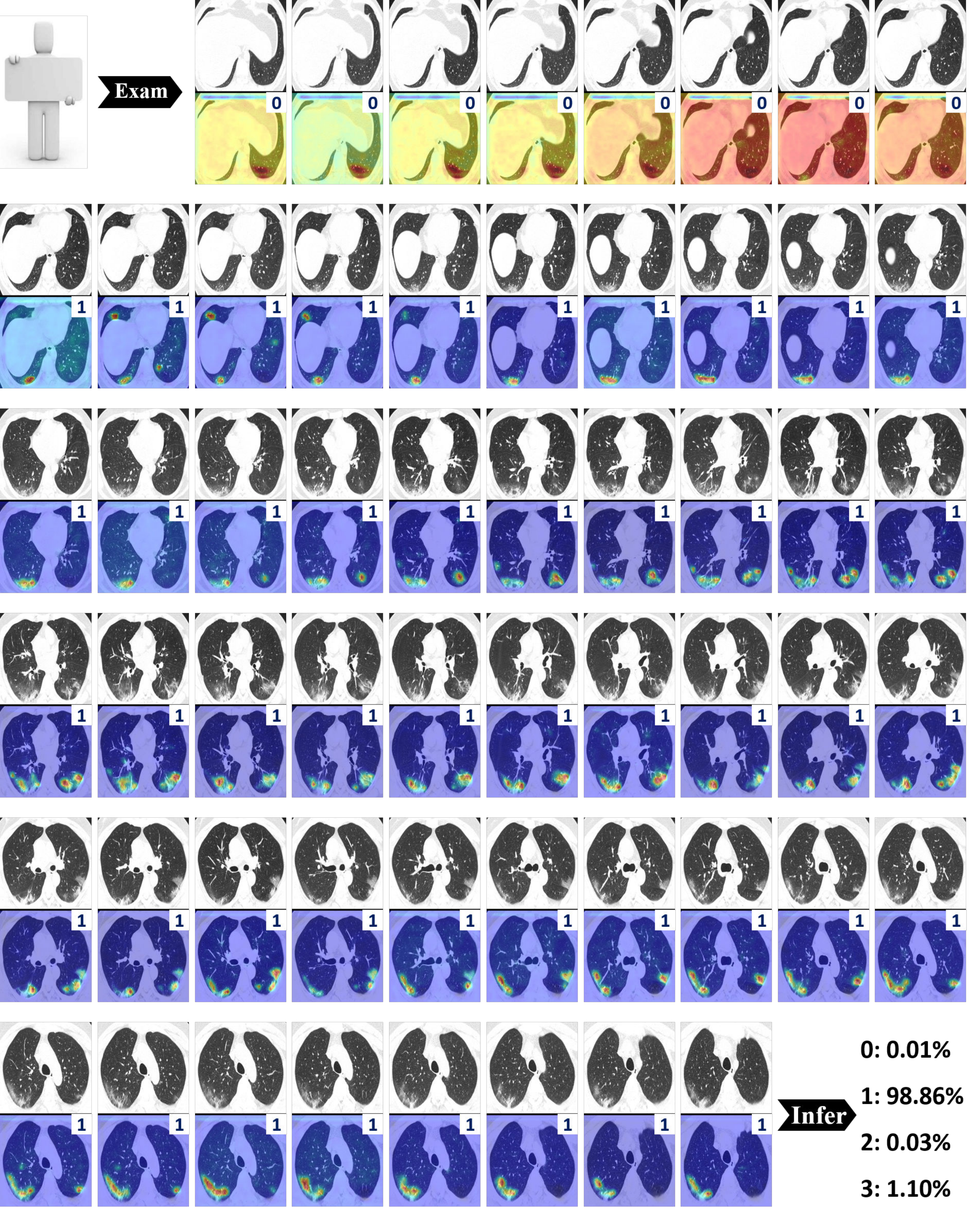} 
\par\end{centering}
\caption{\label{fig:visualization-patient-1} The visualizations of our system outputs for one COVID-19 patient. 
We show a CT sequence by sampling every five slices. 
The lesion location maps, with the predicted slice-level diagnosis on the upper right, are attached at the bottom of the corresponding raw CT slices. 
At the end of the sequence, the probability of each category, predicted by the patient-level classification network is provided, \emph{i.e.}, 0: Healthy; 1: COVID-19; 2: H1N1; 3: CAP. Best viewed in color and zoom in.}
\end{figure}

To fully demonstrate the practicability and effectiveness of our system, we further simulate its real-world application and present the outputs, \emph{i.e.}, the diagnosis assessment of diseases and the localization of lesions. 
Concretely, we randomly select two COVID-19 patients, and feed their CT exams into our system. The full outputs are illustrated in Figure~\ref{fig:visualization-patient-2} and~\ref{fig:visualization-patient-1}. 
Due to page limitation, we show a CT sequence by sampling every five slices. 
The lesion location maps, with the predicted slice-level diagnosis on the upper right, are attached at the bottom of the corresponding raw CT slices. 
At the end of the sequence, the probability of each category predicted by the patient-level classification network is provided (\emph{i.e.}, 0: Healthy; 1: COVID-19; 2: H1N1; 3: CAP). 
As can be seen, our system can accurately locate the lesion areas in each slice, \footnote{For slices that are classified as Healthy, it detects and visualizes the areas of `without lesions' (described in Sec.~\ref{sec:lesion detection}), so almost the entire image is highlighted.} and these areas also have good continuity in sequence, which are very important characteristics to assist the clinician in diagnosing COVID-19.

\section{DISCUSSION}

In this paper, we proposed a multi-task multi-slice deep learning system (M$^{3}$Lung-Sys) to assist the work of clinicians by simultaneously screening multi-class lung pneumonia, \emph{i.e.}, Healthy, COVID-19, H1N1 and CAP, and locating lesions from both slice-level and patient-level CT exams.
Different from previous studies, which incur high hardware, time and data costs to train 3D CNNs, our system divides this procedure into two stages: slice-level classification and patient-level classification. 
We first utilize the slice-level classification network to classify each CT slice. 
An introduced multi-task learning mechanism makes our model diagnose like a radiologist, first checking whether each CT slice contains any abnormality, and then determining what kind of disease it is. 
Both attention maps and coordinate maps are further applied to provide awareness of small lesions and capture location diversity among different types of pneumonia. 
With the improvements provided by these components, our system achieves a remarkable performance of 94.18\% accuracy, 10.66\% false positive error and 0.0\% false negative error. 
Then, to recover the temporal information in CT sequence slices and enable CT volumes screening, we propose a patient-level classification network, taking as input a volume of CT slice features extracted from the slice-level classification network.
Such a network can achieve the feature interaction and aggregation between different slices for patient-level diagnosis. 
Consequently, it further promotes our system by dramatically reducing the cases of false positive and false disease predication and improving the accuracy by 1.1\%.

Above all, from a clinical perspective, our system can perform differential diagnosis with not only a single CT slice, but also a CT volume. 
The combined outputs of risk assessment (predicted diagnosis) and lesion location maps make it more flexible and valuable to clinicians. 
For example, they can easily estimate the percentage of infected lung areas or quickly check the lesions at any time before making a decision. 
As illustrated in Figure \ref{fig:visualization-patient-2} and \ref{fig:visualization-patient-1} , with these lesion location maps, clinicians are able to understand why the deep learning model gives such prediction, not just face a statistic.

Even with the outstanding performance, there are three limitations remaining in our system that need to be improved. 
First of all, it is very easy for clinicians to distinguish COVID-19 from healthy cases. 
However, from Table 2, we find that our system may still misclassify some healthy people. 
We examine failure cases and find that the main challenge lies in the pulsatile artifacts in pulmonary CT imaging. 
Second, our framework, which contains slice-level and patient-level classification networks, is not end-to-end trainable yet. 
Although it just increases the negligible training and testing time, we hope the end-to-end training manner would be more conducive to the learning and combination of spatial and temporal information.
Third, our proposed localization maps accurately show the location of abnormal regions, which are valuable to clinicians in assisting diagnosis. 
However, they still lack the ability to automatically visualize the unique lesions' distributions for each disease.

In the future, we will attempt to tackle the first and third limitation by improving the attention mechanism to enhance the feature representations. 
Besides, developing the technology of coordinate maps may be an optimal option. 
Frankly speaking, the third obstacle is a very challenging and ideal objective, but we will continuously promote research along this line.
As for the second limitation, we are going to polish our method into a unified framework, through such training mechanism, both spatial and temporal information can complement each other.

\section{Conclusion}
In conclusion, we proposed a novel multi-task multi-slice deep learning system (M$^{3}$Lung-Sys) for multi-class lung pneumonia screening from chest CT images. 
Different from previous 3D CNN approaches, which incur a substantial training cost, our system utilizes two 2D CNN networks, \emph{i.e.}, slice-level and patient-level classification networks, to handle the discriminative feature learning from the spatial and temporal domain, respectively. 
With these special designs, our system can not only be trained with much less cost, including time, data and GPU, but can also perform differential diagnosis with either a single CT slice, or a CT volume. 
More importantly, without any pixel-level annotation for training, our system is able to simultaneously output the lesion localization for each CT slice, which is valuable to clinicians for diagnosis, allowing them to understand why our system gives a particular prediction, rather than just being faced with a statistic. 
According to the remarkable experimental results on 292 testing cases with multi-class lung pneumonia (102 healthy people, 96 COVID-19 patients, 41 H1N1 patients and 53 CAP patients), our system has great potential for clinical application.

\appendices
\section{Non-parametric Assessment}
\label{app:Non-parametric Assessment}

Given a volume of CT exam with $N$ slices in sequence, we feed them into our slice-level classification network to obtain two kinds of probabilities for each slice, $P_{lesion}=(P_{l_{i}}^{0},P_{l_{i}}^{1})_{i=1}^{N}$ and $P_{multi-class}=(P_{m_{i}}^{0},P_{m_{i}}^{1},P_{m_{i}}^{2},P_{m_{i}}^{3})_{i=1}^{N}$, where $P_{lesion}$ means the probability whether a CT slice contains any lesion or not, and $P_{multi-class}$ denotes the predicted probability of multi-class pneumonia assessment (\emph{i.e.}, 0: Healthy; 1: COVID-19; 2: H1N1; 3: CAP). Then, we derive the final probabilities of four classes for each slice from $P_{lesion}$ and $P_{multi-class}$, which can be expressed as follows,

\begin{equation}
\footnotesize{
\begin{aligned}
p=(&p_{i}^{0},p_{i}^{1},p_{i}^{2},p_{i}^{3})_{i=1}^{N}, \\
p_{i}^{0}=&P_{l_{i}}^{0}+P_{l_{i}}^{1}\times P_{m_{i}}^{0}, \\
p_{i}^{1}=P_{l_{i}}^{1}\times P_{m_{i}}^{1},~~
p_{i}^{2}&=P_{l_{i}}^{1}\times P_{m_{i}}^{2},~~
p_{i}^{3}=P_{l_{i}}^{1}\times P_{m_{i}}^{3}.
\end{aligned}}
\end{equation}

Intuitively, if all or most of slices are predicted as Health, the patient has a very high chance of being healthy. Otherwise, he will be diagnosed as either COVID-10, H1N1 or CAP, according to CT imaging manifestation. 
To simulate this process, our proposed non-parametric holistic assessment on patient-level can be formulated as follows,

\begin{equation}
\footnotesize{
\left\{
\begin{aligned}
&\text{Healthy}, &if~~&\frac{\mathfrak{N}^{0}}{N}>T \\  
&\text{COVID-19}, &if~~&\frac{\mathfrak{N}^{0}}{N}\leq T,~\max\left(\mathfrak{N}^{1},\mathfrak{N}^{2},\mathfrak{N}^{3}\right)==\mathfrak{N}^{1}\\
&\text{H1N1}, &if~~&\frac{\mathfrak{N}^{0}}{N}\leq T,~\max\left(\mathfrak{N}^{1},\mathfrak{N}^{2},\mathfrak{N}^{3}\right)==\mathfrak{N}^{2}\\
&\text{CAP}, &if~~&\frac{\mathfrak{N}^{0}}{N}\leq T,~\max\left(\mathfrak{N}^{1},\mathfrak{N}^{2},\mathfrak{N}^{3}\right)==\mathfrak{N}^{3} \\
\end{aligned}
\right.
}
\end{equation}

\noindent where $\mathfrak{N}^{k}=\sum_{i=1}^{N}\delta(\max p_{i}-p_{i}^{k}) $ and $\delta(\ast)=1$ if $\ast = 0$, otherwise, $\delta(\ast)=0$. 
$T$ is a hyper-parameter to control the degree of `most of', that is, the proportion of healthy (normal) slices.
Normally, the chest CT slices of healthy people should be nearly or completely all normal. Therefore, without loss of generality, in this paper, we set it as a reasonable and acceptable value, \emph{i.e.}, $T=0.99$.

\section*{Acknowledgment}

The authors would like to thank all of clinicians, patients and researchers who gave valuable time effort and support for this project, especially in data collection and annotation
Additionally, we appreciate the contribution to this paper by Wenxuan Wang (for the help of paper revision), Junlin Hou for her suggestion in 3D baselines) and Longquan Jiang (for the assistance of data processing).

\bibliographystyle{IEEEtran}
\bibliography{mybibfile}

\begin{thebibliography}{10}
\providecommand{\url}[1]{#1}
\csname url@samestyle\endcsname
\providecommand{\newblock}{\relax}
\providecommand{\bibinfo}[2]{#2}
\providecommand{\BIBentrySTDinterwordspacing}{\spaceskip=0pt\relax}
\providecommand{\BIBentryALTinterwordstretchfactor}{4}
\providecommand{\BIBentryALTinterwordspacing}{\spaceskip=\fontdimen2\font plus
\BIBentryALTinterwordstretchfactor\fontdimen3\font minus
  \fontdimen4\font\relax}
\providecommand{\BIBforeignlanguage}[2]{{%
\expandafter\ifx\csname l@#1\endcsname\relax
\typeout{** WARNING: IEEEtran.bst: No hyphenation pattern has been}%
\typeout{** loaded for the language `#1'. Using the pattern for}%
\typeout{** the default language instead.}%
\else
\language=\csname l@#1\endcsname
\fi
#2}}
\providecommand{\BIBdecl}{\relax}
\BIBdecl

\bibitem{adhikari2020epidemiology}
S.~P. Adhikari, S.~Meng, Y.-J. Wu \emph{et~al.}, ``Epidemiology, causes,
  clinical manifestation and diagnosis, prevention and control of coronavirus
  disease (covid-19) during the early outbreak period: a scoping review,''
  \emph{Infectious diseases of poverty}, vol.~9, no.~1, pp. 1--12, 2020.

\bibitem{chan2020familial}
J.~F.-W. Chan, S.~Yuan, K.-H. Kok \emph{et~al.}, ``A familial cluster of
  pneumonia associated with the 2019 novel coronavirus indicating
  person-to-person transmission: a study of a family cluster,'' \emph{The
  Lancet}, vol. 395, no. 10223, pp. 514--523, 2020.

\bibitem{Interim_Guidelines}
NCIRD,
  ``https://www.cdc.gov/coronavirus/2019-ncov/lab/guidelines-clinical-specimens.html,''
  2020.

\bibitem{ai2020correlation}
T.~Ai, Z.~Yang, H.~Hou \emph{et~al.}, ``Correlation of chest ct and rt-pcr
  testing in coronavirus disease 2019 (covid-19) in china: a report of 1014
  cases,'' \emph{Radiology}, p. 200642, 2020.

\bibitem{chung2020ct}
M.~Chung, A.~Bernheim, X.~Mei \emph{et~al.}, ``{CT imaging features of 2019
  novel coronavirus (2019-nCoV)},'' \emph{Radiology}, vol. 295, no.~1, pp.
  202--207, 2020.

\bibitem{pan2020imaging}
Y.~Pan and H.~Guan, ``{Imaging changes in patients with 2019-nCov},''
  \emph{European Radiology}, feb 2020.

\bibitem{fang2020sensitivity}
Y.~Fang, H.~Zhang, J.~Xie, M.~Lin, L.~Ying, P.~Pang, and W.~Ji, ``Sensitivity
  of chest ct for covid-19: comparison to rt-pcr,'' \emph{Radiology}, p.
  200432, 2020.

\bibitem{kanne2020essentials}
J.~P. Kanne, B.~P. Little, J.~H. Chung, B.~M. Elicker, and L.~H. Ketai,
  ``Essentials for radiologists on covid-19: an update—radiology scientific
  expert panel,'' 2020.

\bibitem{liu2020covid}
M.~Liu, W.~Zeng, Y.~Wen, Y.~Zheng, F.~Lv, and K.~Xiao, ``{COVID-19 pneumonia:
  CT findings of 122 patients and differentiation from influenza pneumonia},''
  \emph{European Radiology}, p.~1, 2020.

\bibitem{chen2019pulmonary}
W.~Chen, X.~Xiong, B.~Xie \emph{et~al.}, ``{Pulmonary invasive fungal disease
  and bacterial pneumonia: a comparative study with high-resolution CT},''
  \emph{American journal of translational research}, vol.~11, no.~7, p. 4542,
  2019.

\bibitem{krizhevsky2012imagenet}
A.~Krizhevsky, I.~Sutskever, and G.~E. Hinton, ``Imagenet classification with
  deep convolutional neural networks,'' in \emph{Advances in neural information
  processing systems}, 2012, pp. 1097--1105.

\bibitem{wang2018non}
X.~Wang, R.~Girshick, A.~Gupta, and K.~He, ``Non-local neural networks,'' in
  \emph{Proceedings of the IEEE conference on computer vision and pattern
  recognition}, 2018, pp. 7794--7803.

\bibitem{qian2019leader}
X.~Qian, Y.~Fu, T.~Xiang, Y.-G. Jiang, and X.~Xue, ``Leader-based multi-scale
  attention deep architecture for person re-identification,'' \emph{IEEE
  transactions on pattern analysis and machine intelligence}, vol.~42, no.~2,
  pp. 371--385, 2019.

\bibitem{wang2020fm2u}
W.~Wang, Y.~Fu, X.~Qian, Y.-G. Jiang, Q.~Tian, and X.~Xue, ``Fm2u-net: Face
  morphological multi-branch network for makeup-invariant face verification,''
  in \emph{Proceedings of the IEEE/CVF Conference on Computer Vision and
  Pattern Recognition}, 2020, pp. 5730--5740.

\bibitem{chen2019hybrid}
K.~Chen, J.~Pang, J.~Wang \emph{et~al.}, ``Hybrid task cascade for instance
  segmentation,'' in \emph{IEEE conference on computer vision and pattern
  recognition}, 2019, pp. 4974--4983.

\bibitem{heinsfeld2018identification}
A.~S. Heinsfeld, A.~R. Franco, R.~C. Craddock, A.~Buchweitz, and F.~Meneguzzi,
  ``Identification of autism spectrum disorder using deep learning and the
  abide dataset,'' \emph{NeuroImage: Clinical}, vol.~17, pp. 16--23, 2018.

\bibitem{kong2019classification}
Y.~Kong, J.~Gao, Y.~Xu, Y.~Pan, J.~Wang, and J.~Liu, ``Classification of autism
  spectrum disorder by combining brain connectivity and deep neural network
  classifier,'' \emph{Neurocomputing}, vol. 324, pp. 63--68, 2019.

\bibitem{liu2014early}
S.~Liu, S.~Liu, W.~Cai, S.~Pujol, R.~Kikinis, and D.~Feng, ``Early diagnosis of
  alzheimer's disease with deep learning,'' in \emph{IEEE international
  symposium on biomedical imaging (ISBI)}.\hskip 1em plus 0.5em minus
  0.4em\relax IEEE, 2014, pp. 1015--1018.

\bibitem{ortiz2016ensembles}
A.~Ortiz, J.~Munilla, J.~M. Gorriz, and J.~Ramirez, ``Ensembles of deep
  learning architectures for the early diagnosis of the alzheimer’s
  disease,'' \emph{IJNS}, vol.~26, no.~07, p. 1650025, 2016.

\bibitem{jo2019deep}
T.~Jo, K.~Nho, and A.~J. Saykin, ``Deep learning in alzheimer’s disease:
  diagnostic classification and prognostic prediction using neuroimaging
  data,'' \emph{Frontiers in aging neuroscience}, vol.~11, p. 220, 2019.

\bibitem{albarqouni2016aggnet}
S.~Albarqouni, C.~Baur, F.~Achilles, V.~Belagiannis, S.~Demirci, and N.~Navab,
  ``Aggnet: deep learning from crowds for mitosis detection in breast cancer
  histology images,'' \emph{IEEE transactions on medical imaging}, vol.~35,
  no.~5, pp. 1313--1321, 2016.

\bibitem{bejnordi2017diagnostic}
B.~E. Bejnordi, M.~Veta, P.~J. Van~Diest \emph{et~al.}, ``Diagnostic assessment
  of deep learning algorithms for detection of lymph node metastases in women
  with breast cancer,'' \emph{JAMA}, vol. 318, no.~22, pp. 2199--2210, 2017.

\bibitem{Deng2019}
L.~Deng, S.~Tang, H.~Fu, B.~Wang, and Y.~Zhang, ``{Spatiotemporal Breast Mass
  Detection Network (MD-Net) in 4D DCE-MRI Images},'' in \emph{MICCAI}, 2019,
  pp. 271--279.

\bibitem{sayres2019using}
R.~Sayres, A.~Taly, E.~Rahimy \emph{et~al.}, ``Using a deep learning algorithm
  and integrated gradients explanation to assist grading for diabetic
  retinopathy,'' \emph{Ophthalmology}, vol. 126, no.~4, pp. 552--564, 2019.

\bibitem{Orlando2020}
J.~I. Orlando, H.~Fu, J.~{Barbosa Breda} \emph{et~al.}, ``{REFUGE Challenge: A
  unified framework for evaluating automated methods for glaucoma assessment
  from fundus photographs},'' \emph{Medical Image Analysis}, vol.~59, p.
  101570, jan 2020.

\bibitem{Fu2020_AGE}
\BIBentryALTinterwordspacing
H.~Fu, F.~Li, X.~Sun \emph{et~al.}, ``{AGE challenge: Angle Closure Glaucoma
  Evaluation in Anterior Segment Optical Coherence Tomography},'' \emph{Medical
  Image Analysis}, vol.~66, p. 101798, dec 2020. [Online]. Available:
  \url{http://arxiv.org/abs/2005.02258
  https://linkinghub.elsevier.com/retrieve/pii/S1361841520301626}
\BIBentrySTDinterwordspacing

\bibitem{sun2016computer}
W.~Sun, B.~Zheng, and W.~Qian, ``Computer aided lung cancer diagnosis with deep
  learning algorithms,'' in \emph{Medical imaging 2016: computer-aided
  diagnosis}, vol. 9785.\hskip 1em plus 0.5em minus 0.4em\relax International
  Society for Optics and Photonics, 2016, p. 97850Z.

\bibitem{ardila2019end}
D.~Ardila, A.~P. Kiraly, S.~Bharadwaj \emph{et~al.}, ``End-to-end lung cancer
  screening with three-dimensional deep learning on low-dose chest computed
  tomography,'' \emph{Nature medicine}, vol.~25, no.~6, pp. 954--961, 2019.

\bibitem{zech2018variable}
J.~R. Zech, M.~A. Badgeley, M.~Liu, A.~B. Costa, J.~J. Titano, and E.~K.
  Oermann, ``Variable generalization performance of a deep learning model to
  detect pneumonia in chest radiographs: a cross-sectional study,'' \emph{PLoS
  medicine}, vol.~15, no.~11, 2018.

\bibitem{li2020artificial}
L.~Li, L.~Qin, Z.~Xu \emph{et~al.}, ``{Artificial intelligence distinguishes
  COVID-19 from community acquired pneumonia on chest CT},'' \emph{Radiology},
  p. 200905, 2020.

\bibitem{li2015automatic}
W.~Li, F.~Jia, and Q.~Hu, ``Automatic segmentation of liver tumor in ct images
  with deep convolutional neural networks,'' \emph{Journal of Computer and
  Communications}, vol.~3, no.~11, p. 146, 2015.

\bibitem{guo2019deep}
Z.~Guo, X.~Li, H.~Huang, N.~Guo, and Q.~Li, ``Deep learning-based image
  segmentation on multimodal medical imaging,'' \emph{TRPMS}, vol.~3, no.~2,
  pp. 162--169, 2019.

\bibitem{Zhou2020TMI}
T.~Zhou, H.~Fu, G.~Chen, J.~Shen, J.~Shen, and L.~Shao, ``{Hi-Net:
  Hybrid-fusion Network for Multi-modal MR Image Synthesis},'' \emph{IEEE
  Transactions on Medical Imaging}, 2020.

\bibitem{wang2020deep}
S.~Wang, B.~Kang, J.~Ma \emph{et~al.}, ``A deep learning algorithm using ct
  images to screen for corona virus disease (covid-19),'' \emph{MedRxiv}, 2020.

\bibitem{wang2020fully}
S.~Wang, Y.~Zha, W.~Li \emph{et~al.}, ``A fully automatic deep learning system
  for covid-19 diagnostic and prognostic analysis,'' \emph{European Respiratory
  Journal}, 2020.

\bibitem{medseg}
T.~Sakinis and H.~B. Jenssen, ``http://medicalsegmentation.com/covid19/,''
  2020.

\bibitem{fan2020inf}
D.-P. Fan, T.~Zhou, G.-P. Ji \emph{et~al.}, ``{Inf-Net: Automatic COVID-19 Lung
  Infection Segmentation from CT Scans},'' \emph{IEEE TMI}, 2020.

\bibitem{ma2020towards}
J.~Ma, Y.~Wang, X.~An \emph{et~al.}, ``{Towards Efficient COVID-19 CT
  Annotation: A Benchmark for Lung and Infection Segmentation},'' \emph{arXiv
  preprint arXiv:2004.12537}, 2020.

\bibitem{shi2020review}
F.~Shi, J.~Wang, J.~Shi, Z.~Wu, Q.~Wang, Z.~Tang, K.~He, Y.~Shi, and D.~Shen,
  ``Review of artificial intelligence techniques in imaging data acquisition,
  segmentation and diagnosis for covid-19,'' \emph{IEEE reviews in biomedical
  engineering}, 2020.

\bibitem{shan2020lung}
F.~Shan, Y.~Gao, J.~Wang \emph{et~al.}, ``Lung infection quantification of
  covid-19 in ct images with deep learning,'' \emph{arXiv preprint
  arXiv:2003.04655}, 2020.

\bibitem{mobiny2020radiologist}
A.~Mobiny, P.~A. Cicalese, S.~Zare \emph{et~al.}, ``{Radiologist-Level COVID-19
  Detection Using CT Scans with Detail-Oriented Capsule Networks},''
  \emph{arXiv preprint arXiv:2004.07407}, 2020.

\bibitem{hu2020weakly}
S.~Hu, Y.~Gao, Z.~Niu \emph{et~al.}, ``{Weakly Supervised Deep Learning for
  COVID-19 Infection Detection and Classification from CT Images},''
  \emph{arXiv preprint arXiv:2004.06689}, 2020.

\bibitem{butt2020deep}
C.~Butt, J.~Gill, D.~Chun, and B.~A. Babu, ``Deep learning system to screen
  coronavirus disease 2019 pneumonia,'' \emph{Applied Intelligence}, p.~1,
  2020.

\bibitem{jin2020development}
C.~Jin, W.~Chen, Y.~Cao \emph{et~al.}, ``{Development and Evaluation of an AI
  System for COVID-19 Diagnosis},'' \emph{medRxiv}, 2020.

\bibitem{yang2020deep}
S.~Yang, L.~Jiang, Z.~Cao \emph{et~al.}, ``{Deep learning for detecting corona
  virus disease 2019 (COVID-19) on high-resolution computed tomography: a pilot
  study},'' \emph{Annals of Translational Medicine}, vol.~8, no.~7, 2020.

\bibitem{wu2019deep}
W.~Wu, X.~Li, P.~Du, G.~Lang, M.~Xu, K.~Xu, and L.~Li, ``{A Deep Learning
  System That Generates Quantitative CT Reports for Diagnosing Pulmonary
  Tuberculosis},'' \emph{arXiv preprint arXiv:1910.02285}, 2019.

\bibitem{gozes2020rapid}
O.~Gozes, M.~Frid-Adar, H.~Greenspan \emph{et~al.}, ``{Rapid ai development
  cycle for the coronavirus (covid-19) pandemic: Initial results for automated
  detection \& patient monitoring using deep learning ct image analysis},''
  \emph{arXiv preprint arXiv:2003.05037}, 2020.

\bibitem{9097297}
X.~{Wang}, X.~{Deng}, Q.~{Fu} \emph{et~al.}, ``{A Weakly-supervised Framework
  for COVID-19 Classification and Lesion Localization from Chest CT},''
  \emph{IEEE Transactions on Medical Imaging}, 2020.

\bibitem{brooks1977quantitative}
R.~A. Brooks, ``A quantitative theory of the hounsfield unit and its
  application to dual energy scanning.'' \emph{Journal of computer assisted
  tomography}, vol.~1, no.~4, pp. 487--493, 1977.

\bibitem{shi2020large}
F.~Shi, L.~Xia, F.~Shan \emph{et~al.}, ``Large-scale screening of covid-19 from
  community acquired pneumonia using infection size-aware classification,''
  \emph{arXiv preprint arXiv:2003.09860}, 2020.

\bibitem{makram2000method}
S.~Makram-Ebeid, ``Method and device for automatic segmentation of a digital
  image using a plurality of morphological opening operation,'' Apr.~4 2000, uS
  Patent 6,047,090.

\bibitem{di1999simple}
L.~Di~Stefano and A.~Bulgarelli, ``A simple and efficient connected components
  labeling algorithm,'' in \emph{ICIAP}.\hskip 1em plus 0.5em minus 0.4em\relax
  IEEE, 1999, pp. 322--327.

\bibitem{he2016deep}
K.~He, X.~Zhang, S.~Ren, and J.~Sun, ``Deep residual learning for image
  recognition,'' in \emph{CVPR}, 2016, pp. 770--778.

\bibitem{deng2009imagenet}
J.~Deng, W.~Dong, R.~Socher, L.-J. Li, K.~Li, and L.~Fei-Fei, ``Imagenet: A
  large-scale hierarchical image database,'' in \emph{IEEE conference on
  computer vision and pattern recognition}.\hskip 1em plus 0.5em minus
  0.4em\relax Ieee, 2009, pp. 248--255.

\bibitem{caruana1997multitask}
R.~Caruana, ``Multitask learning,'' \emph{Machine learning}, vol.~28, no.~1,
  pp. 41--75, 1997.

\bibitem{zhou2016learning}
B.~Zhou, A.~Khosla, A.~Lapedriza, A.~Oliva, and A.~Torralba, ``Learning deep
  features for discriminative localization,'' in \emph{IEEE conference on
  computer vision and pattern recognition}, 2016, pp. 2921--2929.

\bibitem{bernheim2020chest}
A.~Bernheim, X.~Mei, M.~Huang \emph{et~al.}, ``{Chest CT findings in
  coronavirus disease-19 (COVID-19): relationship to duration of infection},''
  \emph{Radiology}, p. 200463, 2020.

\bibitem{liu2018intriguing}
R.~Liu, J.~Lehman, P.~Molino \emph{et~al.}, ``An intriguing failing of
  convolutional neural networks and the coordconv solution,'' in \emph{Advances
  in Neural Information Processing Systems}, 2018, pp. 9605--9616.

\bibitem{vaswani2017attention}
A.~Vaswani, N.~Shazeer, N.~Parmar \emph{et~al.}, ``Attention is all you need,''
  in \emph{NIPS}, 2017, pp. 5998--6008.

\bibitem{paszke2017automatic}
A.~Paszke, S.~Gross, S.~Chintala \emph{et~al.}, ``Automatic differentiation in
  pytorch,'' 2017.

\bibitem{chen2019med3d}
S.~Chen, K.~Ma, and Y.~Zheng, ``Med3d: Transfer learning for 3d medical image
  analysis,'' \emph{arXiv preprint arXiv:1904.00625}, 2019.

\bibitem{wang2020weakly}
X.~Wang, X.~Deng, Q.~Fu, Q.~Zhou, J.~Feng, H.~Ma, W.~Liu, and C.~Zheng, ``A
  weakly-supervised framework for covid-19 classification and lesion
  localization from chest ct,'' \emph{TMI}, 2020.

\bibitem{shi2020radiological}
H.~Shi, X.~Han, N.~Jiang \emph{et~al.}, ``{Radiological findings from 81
  patients with COVID-19 pneumonia in Wuhan, China: a descriptive study},''
  \emph{The Lancet Infectious Diseases}, 2020.

\bibitem{song2020emerging}
F.~Song, N.~Shi, F.~Shan \emph{et~al.}, ``{Emerging 2019 novel coronavirus
  (2019-nCoV) pneumonia},'' \emph{Radiology}, vol. 295, no.~1, pp. 210--217,
  2020.

\bibitem{tang2020comparison}
X.~Tang, R.~Du, R.~Wang \emph{et~al.}, ``{Comparison of hospitalized patients
  with acute respiratory distress syndrome caused by covid-19 and H1N1},''
  \emph{Chest}, 2020.

\bibitem{simpson2020radiological}
S.~Simpson, F.~U. Kay \emph{et~al.}, ``{Radiological Society of North America
  Expert Consensus Statement on Reporting Chest CT Findings Related to
  COVID-19. Endorsed by the Society of Thoracic Radiology, the American College
  of Radiology, and RSNA},'' \emph{Radiology: Cardiothoracic Imaging}, vol.~2,
  no.~2, p. e200152, 2020.

\bibitem{kang2012computed}
H.~Kang, K.~S. Lee, Y.~J. Jeong, H.~Y. Lee, K.~I. Kim, and K.~J. Nam,
  ``{Computed tomography findings of influenza A (H1N1) pneumonia in adults:
  pattern analysis and prognostic comparisons},'' \emph{Journal of computer
  assisted tomography}, vol.~36, no.~3, pp. 285--290, 2012.

\bibitem{li2011pneumonia}
P.~Li, D.-J. Su, J.-F. Zhang, X.-D. Xia, H.~Sui, and D.-H. Zhao, ``{Pneumonia
  in novel swine-origin influenza A (H1N1) virus infection: high-resolution CT
  findings},'' \emph{European journal of radiology}, vol.~80, no.~2, pp.
  e146--e152, 2011.

\end{thebibliography}

\end{document}